\documentclass[prd, aps, superscriptaddress, preprintnumbers, twocolumn, floatfix, nofootinbib]{revtex4-2}
\pdfoutput=1

\usepackage{amsfonts}
\usepackage{amsmath}
\usepackage{amssymb}
\usepackage{bm}
\usepackage{dcolumn}
\usepackage{graphicx}   
\usepackage[latin1]{inputenc}
\usepackage{latexsym}
\usepackage{rotating}
\usepackage{hyperref}
\usepackage{graphicx}
\usepackage{color}

\newcommand\be{\begin{equation}}
\newcommand\bea{\begin{eqnarray}}
\newcommand\ee{\end{equation}}
\newcommand\eea{\end{eqnarray}}

\renewcommand{\d}{{\mathrm{d}}}
\renewcommand{\[}{\left[}
\renewcommand{\]}{\right]}
\renewcommand{\(}{\left(}
\renewcommand{\)}{\right)}
\newcommand{\nn}{\nonumber}
\newcommand{\Mpl}{M_{\textrm{Pl}}}

\def\doi{http://doi.org}

\def\d{\mathrm{d}}

\begin{document}

\title{Entanglement Entropy of Cosmological Perturbations}

\author{Suddhasattwa Brahma}
\email{suddhasattwa.brahma@gmail.com}
\affiliation{Department of Physics, McGill University, Montr\'{e}al, QC, H3A 2T8, Canada}

\author{Omar Alaryani}
\email{omar.alaryani@mail.mcgill.ca}
\affiliation{Department of Physics, McGill University, Montr\'{e}al, QC, H3A 2T8, Canada}

\author{Robert Brandenberger}
\email{rhb@physics.mcgill.ca}
\affiliation{Department of Physics, McGill University, Montr\'{e}al, QC, H3A 2T8, Canada}
\affiliation{Institutes of Theoretical Physics and of Particle Physics and Astrophysics, ETH Z\"urich, CH-8093 Z\"urich, Switzerland}

\date{\today}

\begin{abstract}

\noindent We show that the entropy of cosmological perturbations originating as quantum vacuum fluctuations in the very early universe, including the contribution of the leading nonlinear interactions, can be viewed as momentum space entanglement entropy between sub- and super-Hubble modes. The interactions between these modes causes decoherence of the super-Hubble fluctuations which, in turn, leads to a non-vanishing entropy of the reduced density matrix corresponding to the super-Hubble inhomogeneities. In particular, applying this to inflationary cosmology reveals that the entanglement entropy produced by leading order nonlinearities \textit{dominates} over that coming from the squeezing of the vacuum state unless inflation lasts for a very short period. Furthermore, demanding that this entanglement entropy be smaller than the thermal entropy at the beginning of the radiation phase of standard cosmology leads to an upper bound on the duration of inflation which is similar to what is obtained from the Trans-Planckian Censorship Conjecture.

\end{abstract}

\maketitle

\section{Introduction} 
\label{sec:intro}

There has recently been a lot of interest in entanglement entropy in the context of quantum field theory and gravity (see \textit{e.g.}, \cite{Review} for reviews). In particular, the entanglement entropy of a conformal field theory is holographically related to properties of the bulk in the context of the AdS/CFT (anti-de-Sitter bulk/conformal  field theory on the boundary \cite{AdS}) correspondence (see \textit{e.g.}, \cite{holo}). In the same context, entanglement entropy can be related to properties of black holes in the AdS bulk \cite{Malda}. The relationship between the bulk Einstein equations and properties of entanglement of the boundary CFT was explored in \cite{holo2}. Entanglement entropy considerations have also been applied directly to black holes physics (see \cite{Solo} for a review), and to de Sitter space in \cite{Pimentel, Diaz}. There are also attempts to build up space-time itself from quantum entanglement \cite{Mav}. 

Most considerations of entanglement are based on a position space separation of the domain; for example, the separation between the inside of a black hole and the outside. However, in cosmology it is more natural to work in momentum space because it is the properties of the momentum modes of cosmological fluctuations which are generally probed (such as the power spectrum). Momentum space entanglement has been considered in \cite{Bala} (see also \cite{others}), and we will use methods from that work extensively.

Entanglement is a crucial, and rather essential, feature of quantum mechanical systems. In many early universe scenarios, the cosmological fluctuations which we measure today are postulated to emerge from quantum vacuum perturbations. This is the case not only in inflationary cosmology \cite{Mukh}, but also in the {\it Ekpyrotic} scenario \cite{Ekp} and in the {\it matter bounce} scenario \cite{FB}. Cosmological perturbations (see \textit{e.g.}, \cite{MFB, RHBrev} for reviews) are small amplitude fluctuations about the homogeneous and isotropic cosmological background. Because of their small amplitude, the inhomogeneities are generally described in Fourier space. To leading order, each Fourier mode evolves independently, and each mode obeys a harmonic oscillator equation with a time-dependent mass. The Hubble radius $H^{-1}(t)$ (where $H$ is the Hubble expansion rate) plays a key role in the dynamics of the modes: on sub-Hubble scales the canonical fluctuation variable oscillates, while it is squeezed on super-Hubble scales.

Successful early universe scenarios have the common feature that the fluctuation modes which are probed today in cosmological observations were sub-Hubble in the early universe phase, thus allowing a causal generation mechanism. In the classes of models we consider here, the initial state for the fluctuations is taken to be the quantum vacuum state\footnote{String gas cosmology \cite{BV} does not fit into this class since there the initial fluctuations are taken to be thermal.}. When the fluctuation modes exit the Hubble radius, their state becomes a squeezed vacuum state. The Hilbert space of states thus naturally divides into two parts - the super-Hubble mode space ${\cal{H}}_A(t)$ and the sub-Hubble mode space ${\cal{H}}_B(t)$:
\begin{eqnarray}\label{Hilbert_Space}
{\cal{H}}_A(t) \, &=& \, \prod {\cal{H}}_k \,\,\,   |k| < H_c(t) \, \nonumber \\
{\cal{H}}_B(t) \, &=& \, \prod {\cal{H}}_k \,\,\,   |k| \geq H_c(t) \,
\end{eqnarray}
where ${\cal{H}}_k$ is the harmonic oscillator Hilbert space of the k'th mode and $H_c^{-1}(t)$ stands for the comoving Hubble radius.  It is natural to consider the space of super-Hubble modes to be the system we consider, and the space of sub-Hubble modes to be the bath which we integrate over. Note that the comoving Hubble radius decreases as a function of time in the early universe phase of the models which we consider. This means that modes exit the Hubble radius. Hence, the boundary between the two Hilbert spaces ${\cal{H}}_A$ and ${\cal{H}}_B$ depends on time: the dimension of the system Hilbert space is increasing. This is a specific feature of a system on a dynamically expanding background. Furthermore, although not explicitly stated above, we shall assume an ultraviolet (UV) cutoff ($\Mpl$) for the bath modes so that there is always a constant supply of modes which we integrate over. We assume that some underlying UV theory is able to provide the details of the dynamics of the modes lying in the range $k>\Mpl$ and shall not consider them in our work.  

As mentioned above, in this paper, we consider the entropy of the space of super-Hubble modes which results from the entanglement with the bath of sub-Hubble modes. The question of entropy of cosmological perturbations has been considered previously. For example, in \cite{Tom1, Tom2, Tom3} the entropy of a classical field was studied, and the results were applied to compute the entropy of cosmological perturbations and gravitational waves in an inflationary universe. In \cite{Tom1, Tom2, Tom3}, the source of entropy can be traced back to the loss of information about the phases of the fluctuations for super-Hubble modes, while a similar calculation for the coherent state basis was shown in \cite{Matacz:1992tp}. In \cite{GG}, the issue of entropy of cosmological perturbations was reconsidered, taking the loss of information which leads to entropy generation to be the loss of information due to the spreading of the wave function of the super-Hubble modes which results from squeezing. Entropy generation as a consequence of coupling to an environment was studied in \cite{KPS2}. In \cite{CP}, entropy generation of cosmological fluctuations as a consequence of a truncation of the hierarchy of Green's functions was considered.

What was not considered in these previous works on entropy generation is the role of nonlinearities. Because of the nonlinear nature of the Einstein equations, there is always a mixing of modes for cosmological perturbations. In particular, there is a mixing between the sub- and super-Hubble modes. As discussed in \cite{KPS, Martineau, Cliff, Nelson}, this leads to decoherence of the reduced density matrix of super-Hubble modes\footnote{See also \cite{IF} where the decoherence of super-Hubble modes as a consequence of the interaction with sub-Hubble modes was studied using different techniques, \cite{GW} where the decoherence through interaction with gravitational waves was considered, and \cite{Env, Env2} where decoherence due to coupling to a more general environment was analyzed.}. This decoherence is crucial in order to explain why the cosmological perturbations become classical even though they have a quantum origin. The resulting density matrix of the super-Hubble modes is no longer that of a pure state, and hence leads to a non-vanishing entropy which we compute in this paper. We stress that, as shall become apparent later on, we calculate a lower bound on the amount of entanglement entropy of  scalar density perturbations, produced in any model of inflation, due to the minimal gravitational nonlinearities which must always be present. Additional couplings or fields, or considering interactions between scalar and tensor modes, would lead to enhanced amounts of entropy production.

There are some similarities between our work and that of \cite{Tom4}, where decoherence through neglecting observationally inaccessible correlators was considered, and that of \cite{Tom5} where decoherence via entropy field loops was studied (decoherence of fluctuations through entropy loops was considered earlier in \cite{Tom6}). There is also a connection with the work of \cite{Boya} where super-Hubble entanglement through inflaton decay was considered.

Our notation is as follows: We use natural units in which the speed of light, Planck's constant and Boltzmann's constant are set to one. We consider a spatially flat background cosmology such that the metric can be written as
\begin{eqnarray}\label{FLRW}
\d s^2 \, = \, - a^2\({\eta}\) \bigl[\d\eta^2 - \d{\bf{x}}^2 \bigr] \, ,
\end{eqnarray}
where $\eta$ is the conformal time which is related to the physical time $t$ via $dt = a d\eta$, and ${\bf{x}}$ are the comoving spatial coordinates. The Hubble parameter is given in terms of the scale factor $a(t)$ by
\begin{eqnarray}
H(t) \, = \, \frac{\dot{a}}{a} \, ,
\end{eqnarray}
where the overdot represents the derivative with respect to $t$. We emphasize that the Hubble radius plays a crucial role in our analysis. Sub-Hubble modes of the canonical fluctuation variable oscillate while those on super-Hubble scales are squeezed \cite{MFB, RHBrev}. We denote the Planck mass by $\Mpl$.

In the next section, we give a first pass at arriving at the entropy of cosmological perturbations due to the squeezing of super-Hubble modes during inflation. In Sec-\ref{analysis}, we review the well-known argument that interaction between the perturbation modes, arising from minimal gravitational nonlinearities, leads to a suppression of the off-diagonal terms in the density matrix for the super-Hubble modes. This justifies an assumption used in Sec-\ref{review} for calculating the entropy due to the squeezed state. Finally, having set up our dominant interaction term in Sec-\ref{analysis}, we go on to calculate the entanglement entropy density for our system (super-Hubble) modes in Sec-\ref{Ent_Entropy}. We estimate an order of magnitude for the upper bound of this quantity and show that it is greater than the entropy for the squeezed vacuum, as calculated in Sec-\ref{review}. In Sec-\ref{TCC}, interestingly we find an  upper bound on the duration of inflation by requiring that this entanglement entropy remains smaller than the thermal entropy produced at the end of inflation\footnote{This bound is similar to the bound obtained \cite{TCC2} by invoking the {\it Trans-Planckian Censorship Conjecture} (TCC) \cite{TCC}.}.  We discuss our main findings in Sec-\ref{conclusion}.

\section{Reduced Density Matrix of Super-Hubble Modes}\label{review}

\subsection{The Squeezed Vacuum}
We consider linear scalar cosmological perturbations about the background metric \eqref{FLRW}. Assuming that the matter source of the fluctuations has no anisotropic stress, the perturbations are described by a single field $\zeta(x, t)$, the curvature perturbation in comoving gauge. The metric including these fluctuations is
\begin{eqnarray}
\d s^2 \, = \, - a^2(\eta) \bigl[\d\eta^2 - (1 + 2 \zeta) \d{\bf{x}}^2 \bigr] \, .
\end{eqnarray}
The action for cosmological perturbations has a canonical kinetic term if we use the rescaled field (we are following the notation of \cite{Shandera:2017qkg})
\begin{eqnarray}
\chi(x, \eta) \, \equiv \, z(\eta) \zeta(x, \eta) 
\end{eqnarray}
with
\begin{eqnarray}
z^2(\eta) \, \equiv \, 2\, \epsilon_H\, a^2\, M_{pl}^2 \,c_s^{-2} \,, 
\end{eqnarray}
where $\epsilon_H$ is the first ``slow-roll'' parameter defined via
\begin{eqnarray}
\epsilon_H \, \equiv \, - \frac{\dot{H}}{H^2} \, ,
\end{eqnarray}
and $c_s^2$ is the speed of sound squared of the matter source. Although, later on, we shall only consider models of single-field inflation with no derivative self-couplings, we are keeping $c_s \neq 1$ at this stage so that our expressions remain as general as possible\footnote{In the case $c_s = 1$, the action is $\int \d^4x\; \frac{1}{2} \bigr[ \(\partial_{\mu} \chi\)^2 - \frac{z^{\prime \prime}}{z} \chi^2 \bigr]$.}.

The linear cosmological perturbations about the classical background geometry can be canonically quantized \cite{Mukh}. We insert the ansatz for the fluctuating metric and matter into the total action (joint gravitational and matter action) and expand to quadratic order. Since at linear order each Fourier mode evolves independently, we can reduce the quantization to the standard quantization of a set of harmonic oscillators, the oscillators having a time dependent mass coming from the time dependence of the background. In terms of the usual ladder operators, the quadratic Hamiltonian $H_2$ corresponding to scalar cosmological perturbations takes the form
\begin{eqnarray}\label{H_2}
H_2 \, &=& \, \frac{1}{2} \int \frac{\d^3 k}{\(2\pi\)^3} \left[  c_s k \(c_\textbf{k} c^\dagger_\textbf{k} + c_\textbf{-k} c^\dagger_\textbf{-k}\)  \right] \nonumber \\
&-& \,  \frac{1}{2} \int \frac{\d^3 k}{\(2\pi\)^3} \left[  i \(\frac{z'}{z}\) \(c_\textbf{k} c_\textbf{-k}-c^\dagger_\textbf{k} c^\dagger_\textbf{-k}\) \right] \,,
\end{eqnarray}
where a prime denotes a derivative with respect to conformal time. As can be seen from \eqref{H_2}, the squeezing term dominates in the limit $aH \gg c_sk$, for a given mode. In other words, the time-dependent squeezing interaction is dominant for super-Hubble modes.  

This quadratic Hamiltonian generates the following equation of motion for the ladder operators
\begin{eqnarray}
	\frac{\d c_\textbf{k}}{\d \eta} \, = \, \(\frac{z'}{z}\) c^\dagger_\textbf{k} - i c_s k c_\textbf{k}\,.
\end{eqnarray}
Given an initial condition at an instant of time, $\eta_0$, we can solve for this as
\begin{eqnarray}
c_\textbf{k}\(\eta\) \, &=& \,  e^{i\theta_k\(\eta\)} \cosh\[r_k\(\eta\)\] c_\textbf{k}\(\eta_0\) \nonumber \\
&+& \, e^{-i\theta_k\(\eta\) + 2i\phi_k\(\eta\)} \sinh\[r_k\(\eta\)\] c^\dagger_\textbf{-k}\(\eta_0\)\,.
\end{eqnarray}
In the above, $r_k$ and $\phi_k$ are the squeezing parameter and the squeezing angle, whereas $\theta_k$ denotes the action of the rotation operator. The number of particles in a given mode $k$ is proportional to the squeezing parameter $n_k \sim \sinh^2r_k$. For inflation, the leading order time-dependence of these parameters is given by \cite{Albrecht:1992kf}
\begin{eqnarray}\label{squeezing_paramters}
	r_k\(\eta\) &=& -\sinh^{-1}\(\frac{1}{2c_s k\eta}\)\,,\label{r_k}\\
	\phi_k\(\eta\) &=& - \frac{\pi}{4} - \frac{1}{2} \tan^{-1}\(\frac{1}{2c_s k\eta}\)\,,\\
	\theta_k\(\eta\) &=& -k\eta - \tan^{-1}\(\frac{1}{2c_s k\eta}\)\,.
\end{eqnarray}

Given the quadratic Hamiltonian, the evolution operator $U_0(\eta)$ can be written as 
\begin{eqnarray}
	U_0\(\eta, \eta_0\) \left|0_\textbf{k}, 0_\textbf{-k}\right\rangle = S_k\(\eta\) R_k\(\eta\) \left|0_\textbf{k}, 0_\textbf{-k}\right\rangle\,,
\end{eqnarray}
where $S_k\(r_k,\phi_k\)$ and $R_k\(\theta_k\)$ are the two-mode squeezing and rotation operators, respectively, which are defined as \cite{Albrecht:1992kf}
\begin{eqnarray}
	S_k &:=& \exp\[\dfrac{r_k}{2}\(e^{-2i\phi_k}\,c_{-{\bf k}} c_{\bf k} - {\rm h.c.}\)\]\,,\\
	R_k &:=& \exp\[-i\theta_k \(c^\dagger_{{\bf k}} c_{\bf k} +c^\dagger_{-{\bf k}} c_{-{\bf k}} +1\)\]\,.
\end{eqnarray}
At the level of the quadratic Hamiltonian, the $U_0(\eta)$ is unitary. However, once interaction terms are introduced, the evolution becomes necessarily non-unitary in the presence of bath modes \cite{Shandera:2017qkg}. The effect of the rotation operator is only to change the phase and would be of no consequence to us, and hence we drop it from hereon. The effect of the two-mode squeezing operator on the vacuum leads to the squeezed vacuum, which is defined as
\begin{eqnarray}\label{2modeSqueezed}
	 \left|SQ\(k,\eta\) \right\rangle \, 
	&\equiv & \,  S_k\(r_k,\phi_k\) \left|0_\textbf{k}, 0_\textbf{-k}\right\rangle  \\
	&=& \, \frac{1}{\cosh r_k}\sum_{n=0}^{\infty} e^{-2in\phi_k} \tanh^nr_k \left|n_\textbf{k}, n_\textbf{-k}
	 \right\rangle\, , \nonumber
\end{eqnarray}
where  
\begin{eqnarray}
\left|n_\textbf{k}, n_\textbf{-k} \right\rangle \, \equiv \, \[\frac{1}{n!}\(c^\dagger_\textbf{k} c^\dagger_\textbf{-k}\)^n\] \left|0_\textbf{k}, 0_\textbf{-k}\right\rangle \, .
\end{eqnarray}
For a given mode $k$, it is easy to see that this state is normalized, as follows:
\begin{eqnarray}\label{Normalization}
&& \left\langle SQ\(k,\eta\) | SQ\(k,\eta\) \right\rangle \, \nonumber \\
&& = \, \frac{1}{\cosh^2 r_k} \sum_{n=0}^{\infty} \sum_{m=0}^{\infty} e^{-2i(n-m)\phi_k} \tanh^{(m+n)} r_k\, \delta_{m,n}\nn\\
&& = \, \frac{1}{\cosh^2 r_k} \sum_{n=0}^{\infty} \tanh^{2n} r_k = 1\,,
\end{eqnarray}
as required. The squeezed vacuum of all the modes can be obtained in a straightforward manner as the tensor product state
\begin{eqnarray}\label{Squeezed}
	\left|SQ(\eta)\right\rangle \, \equiv \,  \prod_k 	\left|SQ(k, \eta)\right\rangle\,.
\end{eqnarray}

\subsection{The Reduced Density Matrix}
The straightforward definition of the density matrix, corresponding to the squeezed state given in \eqref{Squeezed}, is 
\begin{eqnarray}
	\rho \, = \,  \left|SQ(\eta)\right\rangle \left\langle SQ(\eta) \right|\,.
\end{eqnarray}
If we calculate the entropy corresponding to this state, naturally this is going to be zero since it is a pure state, given by the evolution of the vacuum under the quadratic Hamiltonian \eqref{H_2}.  More concretely, the density matrix expressed in terms of the two-mode occupation number basis reads
\begin{widetext}
\begin{eqnarray}\label{DensityMatrix}
\rho = \prod_k \prod_p \sum_{n=0}^{\infty} \sum_{m=0}^{\infty}  \frac{1}{\cosh r_k \cosh r_p} e^{-2i\phi_k \(n-m\)}\, \tanh^n r_k \tanh^m r_p \left|n_\textbf{k}, n_\textbf{-k} \right\rangle \left \langle m_\textbf{p}, m_\textbf{-p} \right|\,,
\end{eqnarray}
\end{widetext}
which is still a pure density matrix.

Let us show this more explicitly, as follows. Our state can be written as a product state
\begin{eqnarray}
\left| \psi \right\rangle \, = \, \left| \psi \right\rangle_A \, \otimes \, \left| \psi \right\rangle_B \, ,
\end{eqnarray}
where $\left| \psi \right\rangle_A$ is the product state of all the super-Hubble modes, and $\left| \psi \right\rangle_B$ over the sub-Hubble modes. Since we are focusing on the super-Hubble modes, our reduced density matrix is obtained by tracing over the the sub-Hubble mode Hilbert space.
\begin{eqnarray}
\rho_A \, \equiv \, {\rm{Tr}}_B \rho \, 
= \,  \sum_j \left\langle j  \left| \psi \right\rangle \left\langle \psi \right| j \right\rangle \, , 
\end{eqnarray}
where the sum is over the basis states of the Hilbert space of sub-Hubble modes. In the absence of entanglement between the sub- and super-Hubble modes, and given that the states of both subsystems are pure, the reduced density matrix $\rho_A$ also corresponds to that of a pure state and hence has vanishing entropy.

So far, however, we have neglected any coarse graining or nonlinear effects. In particular, we have neglected entanglement effects between sub- and super-Hubble modes which are inevitably present because the equations of gravity are nonlinear. In the following we will take a first look at the entropy of cosmological perturbations after loss of some information about the state. In the following section we then show that this loss of information is an inevitable consequence of the entanglement between sub- and super-Hubble modes.

\subsection{First View on Entanglement Entropy of Cosmological Pertubations}

In order to get a non-vanishing von-Neumann entropy of the reduced density matrix $\rho_A$, we need to coarse-grain it in a suitable way to derive a mixed density matrix. In \cite{Tom1, Tom2}, it was observed that the phase associated with the squeezing angle is sensitively dependent on the density perturbation whereas the amplitude is not. As a consequence, the coarse-grained entropy in \cite{Tom1, Tom2} was defined by averaging over the squeezing angle, which also leads to decoherence. In our setup, a similar ``averaging'' over the squeezing angle, provided there is a stochastic part to it in addition to what is given in \eqref{squeezing_paramters}, would lead to setting the off-diagonal elements to zero in the number basis, leading to a reduced density matrix of the form
\begin{eqnarray}\label{DensityReduced}
  \rho_\text{sq} =
  \,\,\, \prod_{k} \sum_{n=0}^\infty \frac{1}{\cosh^2\(r_k\)} \tanh^{2n}\(r_k\)  \left|n_\textbf{k}, n_\textbf{-k} \right\rangle \left \langle n_\textbf{k}, n_\textbf{-k} \right|\,.\nn\\
\end{eqnarray}
A different perspective of arriving at the above form for the reduced density matrix would be to consider only the diagonal entries of \eqref{DensityMatrix}, whereas assuming that the off-diagonal elements quickly fall-off to zero. The usefulness of this perspective lies in the fact that one does not have to refer to the phase in order to derive the reduced density matrix. However, now we need to justify our choice of ignoring the off-diagonal elements for the density matrix. One way to argue would be to consider that there are a lot of particles created for a given mode, with opposite momenta, with $\phi_k$ being the phase of each of these particle pairs. Further assuming that these phases contain a random part, one can use the destructive interference, due to group averaging these phases, as being responsible for suppressing the off-diagonal terms. However, in this case we are back to our previous argument of using an averaging procedure over random part of the phases. Instead, one might follow the arguments of \cite{GG, Gasperini:1995yd} to justify the reduction of the density matrix as a result of assuming a distribution of coherent states as our initial state -- instead of the usual vacuum -- as a manifestation of our ignorance regarding initial conditions. If one assumes this as the starting point, it can be shown that the off-diagonal terms are naturally suppressed as long as one invokes equipartition of probabilities for the initial states in the ensemble \cite{Gasperini:1995yd}. We are neither advising this approach nor suggesting that it is better than considering the averaging procedure over random phases, but just pointing out that there have been different justifications for considering the above form of the reduced density matrix \eqref{DensityReduced}. For now, we simply assume our coarse-graining procedure to be one that results in the suppression of the off-diagonal terms, as is common in the literature to calculate the squeezing entropy \cite{Tom3}. In the next section, we will give an improved analysis and explain the decay of the off-diagonal elements as a consequence of decoherence resulting from entanglement between the modes and use that as a heuristic argument for our procedure adopted here.

The von-Neumann entropy associated with this reduced density matrix is given by
\begin{widetext}
\begin{eqnarray}\label{Entropy}
	s^c_{\rm sq} &=& -\text{Tr}\(\rho_\text{sq} \ln \rho_\text{sq}\) \nn\\
	&=& - \prod_{k} \frac{1}{\cosh^2 r_k} \ln\(\prod_{p}\frac{1}{\cosh^2 r_p}\) - \frac{\tanh^2 r_k}{\cosh^2 r_k} \ln\(\prod_{p}\frac{\tanh^2 r_p}{\cosh^2 r_p}\) - \frac{\tanh^4 r_k}{\cosh^2 r_k} \ln\(\prod_{p}\frac{\tanh^4 r_p}{\cosh^2 r_p}\) -\ldots\nn\\
	&=& -\sum_{n=0}^{\infty} \[\prod_{k} \frac{\tanh^{2n}r_k}{\cosh^2 r_k}\; \ln\(\prod_{p} \frac{\tanh^{2n}r_p}{\cosh^2 r_p}\)\]
\end{eqnarray}
First, we expand the product in the logarithm as a sum of logs, i.e. 
\begin{eqnarray}
\ln\(\prod_{k} \frac{\tanh^{2n}r_k}{\cosh^2 r_k}\) = \sum_k \ln\(\frac{\tanh^{2n}r_k}{\cosh^2 r_k}\)\,.
\end{eqnarray}
Using this in \eqref{Entropy}, we can rewrite the entropy density (per comoving volume) as
\begin{eqnarray}
s^c_{\rm sq} = -\(\sum_{n=0}^{\infty} \prod_{p} \frac{\tanh^{2n}r_p}{\cosh^2 r_p} \) \(\sum_{k} \sum_{m=0}^{\infty} \frac{\tanh^{2n}r_k}{\cosh^2 r_k}\; \ln \(\frac{\tanh^{2n}r_k}{\cosh^2 r_k}\) \)\,.
\end{eqnarray}
Using the normalization \eqref{Normalization}, the term in the first parentheses is equal to 1. The entropy gets simplified to
\begin{eqnarray}\label{Squeezed_entropy}
	s^c_{\rm sq} &=& \sum_{k} \sum_{n=0}^{\infty} \frac{\ln \(\cosh^2 r_k \(\tanh r_k\)^{-2n}\)}{\cosh^2 r_k}  \tanh^{2n} r_k \nn\\
	&=& \sum_{k} \frac{\ln \(\cosh^2 r_k \)}{\cosh^2 r_k} \sum_{n=0}^{\infty} \tanh^{2n} r_k - \sum_{k} \frac{\ln \(\tanh^2 r_k \)}{\cosh^2 r_k} \sum_{n=0}^{\infty} \[n \tanh^{2n}r_k\]\nn\\
	&=& \sum_{k} \ln \(1+\sinh^2 r_k\) - \sum_{k} \sinh^2 r_k \ln\(\tanh^2 r_k\)\nn\\
	&=& \sum_k \[\(1+\sinh^2 r_k\) \ln \(1+\sinh^2 r_k\) - \sinh^2 r_k \ln \(\sinh^2 r_k\)\]\,.
\end{eqnarray}
\end{widetext}
In the large occupation number limit, $n_k = \sinh^2 r_k \gg 1$, we get back the same expression for the entropy density $s \approx \sum_{k} \ln \(\sinh^2 r_k\)$, as derived in \cite{Tom1, Tom2}. However, we derived this result from the von-Neumann entropy formula for a quantum density matrix instead of using the Shannon entropy for a classical field. Note that one should expect that our expression matches that for the classical calculation, done earlier, only in the large squeezing limit. In this sense, one should view $\sum_{k} \ln \(\sinh^2 r_k\)$ as the classical limit of the von-Neumann entropy calculated here within a quantum field theoretic approach, and it is thus compatible with previous results \cite{Tom1, Tom2} of considering the entropy of a classical field. Our result also matches with previous works as presented in \cite{GG}. 

In the case of slow-roll inflation with an approximately constant Hubble constant we can estimate the resulting entropy density by integrating over all super-Hubble modes and apply a infrared cutoff: we do not consider modes with wavelengths larger than the Hubble radius $H^{-1}$ at the beginning of inflation. With the convention that the scale factor is set to one at the beginning of inflation, this implies that in (\ref{Squeezed_entropy}) we need to integrate over all values of $k$ with $H < k < aH$. At any time, this integral is dominated by the modes exiting the Hubble radius at that time, and we thus obtain\footnote{A more explicit calculation for this has been shown in Sec-\ref{Ent_Entropy}.}
\begin{eqnarray}
s^c_{\rm sq} \, \sim  \, a^3 H^3 \, .
\end{eqnarray}
To obtain the entropy density per physical volume element, we have to divide the above by $a^3$, and we hence get
\begin{eqnarray}\label{Entropy_Squeezed_estimate}
s_{\rm sq} \, \sim \, H^3 \, .
\end{eqnarray}

Before moving on, let us note that the entropy calculated in this section is not quite an entanglement entropy as it arises from the squeezing of the cosmological perturbations. The way we manage to get a nonzero result for a density matrix arising from a quadratic Hamiltonian \eqref{H_2} is by employing some yet-to-be-specified coarse-graining, due to which the pure density matrix in \eqref{DensityMatrix} is reduced to a mixed one \eqref{DensityReduced}, by ignoring the off-diagonal terms. In the next section, we shall give a more nuanced argument as to how gravitational nonlinearities, responsible for decohering the quantum fluctuations into classical perturbations, necessarily render the density matrix diagonal. Using this result, we shall argue that the diagonalization adopted in this section is a well-motivated one and is the reason why it correctly reproduces the entropy of the squeezed vacuum. In this way, the entanglement between sub- and super-Hubble modes, due to mode-mixing arising from gravitational non-linearities, is also indirectly responsible, albeit by providing a heuristic justification for our coarse-graning procedure, for the entropy of cosmological perturbations calculated above\footnote{The key point is that this is in addition to the explicit entanglement entropy due to such interaction terms which we shall calculate later on.}.

\section{Nonlinearities, Decoherence and Entropy Generation}\label{analysis}

Here we review the analysis of \cite{Nelson} which shows how the purely gravitational interactions which are inevitably present because of the nonlinearity of General Relativity lead to a decoherence of the reduced density matrix of the super-Hubble modes as a consequence of the interaction with the sub-Hubble fluctuations. For our purposes, we will focus on the case of inflation.

We shall now take into account the effects of the cubic Hamiltonian in addition to the quadratic Hamiltonian discussed in the previous section. This is the leading term which generates entanglement between the sub- and super-Hubble modes. We are considering the full cubic action for the density perturbations in the presence of a single matter field. If the matter is a canonically normalized scalar field, then the speed of sound $c_s = 1$. In more general models, $c_s^2$ can be smaller than one, and this can significantly increase the size of the cubic interaction terms, resulting in a significant contribution to the equilateral-shape non-Gaussianity parameter $f_{NL}$. However, as a first pass, let us only consider vanilla matter models with $c_s=1$, which should be sufficient to estimate a lower bound on the entanglement entropy for models of inflation. 

We take the form of the cubic contribution to the Hamiltonian  from \cite{Adshead:2011bw}, from now on setting $c_s=1$, which is a generalization of the results from \cite{Maldacena:2002vr}.
\begin{align}\label{CubicLag}
S_3=\Mpl^2 \int \d t\, \d^3 x &\left[a^3\epsilon_H^2 \zeta\dot{\zeta}^2+a\epsilon_H^2\zeta(\partial\zeta)^2- 2a\epsilon_H\dot{\zeta}\partial_i\zeta\partial_i{\tilde{\chi}} \right. \nonumber \\
& \left. +a^3\epsilon_H(\dot{\epsilon}_H-\dot{\eta}_H)\mathcal{\zeta}^2\dot{\mathcal{\zeta}}+ \frac{\epsilon^2_H}{2}a\partial_i\zeta\partial_i{\tilde{\chi}}\right.\nonumber\\
&\left.-\frac{\d}{\d t}\left(a^3\epsilon_H(\epsilon_H-\eta_H)\mathcal{\zeta}^2\dot{\mathcal{\zeta}}\right)\right]\,
\end{align}
where ${\tilde{\chi}}=a^2\epsilon_H\partial^{-2}\dot{\zeta}$. We have also introduced the second ``slow-roll'' parameter:
\begin{eqnarray}
\eta_H \, = \frac{1}{H}\frac{\dot\epsilon_H}{\epsilon_H}\,.
\end{eqnarray}

We shall ignore the non-local terms that contain ${\tilde{\chi}}$ since those are not the dominant terms in the action. Additionally, there are also terms which would get cancelled with each other (such as the $\dot{\eta_H}$ term in the second line would get cancelled by a similar term from the third line of \eqref{CubicLag}). Since, in the case of inflation, the dominant mode of $\zeta$ has frozen out on super-Hubble scales, we will neglect interaction terms which contain ${\dot{\zeta}}$. Furthermore, we shall restrict our analyses only to the leading order terms in the slow-roll parameters, and would thus be left with the second term in the first line of \eqref{CubicLag} (the other terms being higher orders in $\epsilon_H$ and $\eta_H$, or contain a $\dot{\zeta}$). Hence, the dominant term in the interaction Hamiltonian is (after integration by parts, and recalling that $H_{\rm int} = - \mathcal{L}_{\rm int}$) 
\begin{eqnarray}\label{Hint1}
H_{\rm int} \, = \, \frac{\Mpl^2}{2}\int d^3x \; \epsilon_H^2\, a \,\zeta^2 (\partial^2 \zeta) \, .
\end{eqnarray}
The $H_{\rm int}$ we are considering arises purely from gravitational non-linearities, originating from the cubic Lagrangian given in \eqref{CubicLag}. As discussed above, in a model of single-field slow roll inflation without any derivative self-interaction, this would be the dominant term. However, for a nontrivial speed of sound model, there can a different term which significantly enhances the cubic interaction. This would lead to both a faster rate of decoherence as well as a greater amount of entanglement entropy. In this sense, our calculation should be understood to yield the minimum amount of entanglement entropy that \textit{must} be produced in any inflationary model; multiple fields or more complicated interactions would only enhance our results.

Note here that there is an additional term, not shown above in the cubic Lagrangian, that is of the exact same form, $\zeta^2 (\partial^2 \zeta)$, but with a pre-factor $\epsilon_H \eta_H a$ \cite{Chen:2006nt}. This term is part of a large number of terms which are typically removed by a field redefinition \cite{Maldacena:2002vr} and do not affect the correlation functions for calculating the bispectrum. Strictly speaking, we should keep this term if we are interested in calculating the entropy corresponding to the $\zeta$ field (and not  for the redefined one). However, we drop it here to avoid additional clutter since it is straightforward to include its effects at the end by adding a factor of $\epsilon_H\eta_H$, in addition to the $\epsilon_H^2$ in \eqref{Hint1}, to our results. 

Having setup our interaction terms, we begin the evolution at the conformal time $\eta_0$, in a pure Gaussian product state of all of the modes, which has the wave function
\begin{eqnarray}
\Psi[A, B] (\eta_0) \, = \, \Psi_G[A] (\eta_0) \Psi_G[B] (\eta_0) \, ,
\end{eqnarray}
where in this case we have indicated which variables the individual states depend on. As a consequence of the interactions, the state evolves into
\begin{eqnarray}
\Psi[A, B] (\eta) \, = \, \Psi_G[A] (\eta) \Psi_G[B] (\eta) \Psi_I[A, B] (\eta) \, 
\end{eqnarray}
at a later time $\eta$, where the third factor is a consequence of the interaction Lagrangian.

The interaction contribution to the wave function is given by
\begin{eqnarray}
\Psi_I[A, B] (\eta) \, = {\rm{exp}} \bigl[  \int_{k, k', q} \zeta_k \zeta_{k'} \zeta_q {\cal{F}}(k, k', q; \eta) \bigr] \, ,
\end{eqnarray}
where $k, k'$ stand for sub-Hubble modes, and $q$ stands for a super-Hubble mode, and the kernel function ${\cal{F}}(k, k', q; \eta)$ is given by an integration over time of the interaction Hamiltonian in momentum space (see \cite{Nelson} for details) with the property that its imaginary part blows up as $\eta \rightarrow 0$. In the above, the integration runs over all momenta with the property that $k + k' + q = 0$ (momentum conservation).

The reduced density matrix of the super-Hubble modes can be obtained by integrating over the sub-Hubble ones. In the field representation we have
\begin{eqnarray}\label{Red1}
\rho_A(\zeta, {\bar{\zeta}}) \, = \, \int {\cal{D}}_B \Psi[\zeta, B]) \Psi^{*}[{\bar{\zeta}}, B] \, ,
\end{eqnarray}
where ${\cal{D}}_B$ stands for the integration over the sub-Hubble modes $B$. Eq. (\ref{Red1}) yields
\begin{widetext}
\begin{eqnarray}
\rho_A(\zeta, {\bar{\zeta}}) \, &=& \, \Psi_G[\zeta] \psi_G[{\bar{\zeta}}]
\int {\cal{D}}_B |\Psi_G[B]|^2 {\rm{exp}} \bigl[ \int_{k, k', q} \zeta_k \zeta_{k'} \bigl( \zeta_q {\cal{F}}(k, k', q) + {\bar{\zeta}}_q {\cal{F}}^{*}(k, k', q) \bigr)  \bigr] \, \nonumber \\
&\equiv& \, \Psi_G[\zeta] \Psi_G[{\bar{\zeta}}] D[\zeta, {\bar{\zeta}}] \, ,
\end{eqnarray}
where $D[\zeta, {\bar{\zeta}}]$ is the decoherence factor. Focusing on a single super-Hubble mode $q$, the decoherence factor is
\begin{eqnarray}\label{Red2}
D[\zeta, {\bar{\zeta}}] \, \sim \, {\rm{exp}} \bigl[ - \frac{4 \pi (\Delta \zeta_q)^2}{q^3} \int_{k + k' = -q} P_G(k) P_G(k') ({\rm{Im}} {\cal{F}}(k, k', q))^2 \bigr] \, ,
\end{eqnarray}
\end{widetext}
where the time dependence of the factors has been suppressed, where $P_G$ is a property of the Gaussian wavefunction, and 
\begin{equation}
\Delta \zeta_q \, = \, \zeta_q - {\bar{\zeta}}_q \, .
\end{equation}
As is clear from (\ref{Red2}), the decoherence factor decays in time on super-Hubble scales since the imaginary part of ${\cal{F}}$ blows up. Note that the decoherence effect is dominated by the Hubble scale modes. There is no UV divergence in the loop diagram which produces the interaction. This is a consequence of the specific form of our interaction Lagrangian.

To conclude this section, we have reviewed how the interaction with the sub-Hubble modes leads to decoherence of the super-Hubble ones. For a particular mode, decoherence happens after Hubble radius crossing. The important thing for us is the fact that decoherence leads to the damping of the off-diagonal terms of the decoherence functional, such that the reduced density matrix of the super-Hubble modes become diagonal very quickly. The bottom line which we wish to emphasize  is quite common for decoherence during inflation -- time evolution of the density matrix of the system, on interaction with the bath degrees of freedom, leads to a suppression of its off-diagonal terms \cite{Martineau, Martin}. Of course, this ``dynamical diagonalization'' happens when to the density matrix written in terms of the basis of the interaction term. Instead of the Schr\"odinger wave functional approach taken here, one can also demonstrate this by solving the master equation in the Mukhanov-Sasaki variable basis \cite{Martineau}. Let us note that this is not a new calculation which we present here; rather, it is a review of well-established results that the density matrix of the system modes diagonalizes due to interactions with the bath modes, in a basis picked by the interaction term. 

How is then this related to our calculations in the previous section, in which we had calculated the entropy corresponding to the squeezing of the super-Hubble modes, \textit{assuming} that their density matrix turns diagonal? We now give a better justification for choosing to keep only the diagonal elements in our coarse-graining procedure adopted earlier. When interactions between system and bath modes are turned on, then the resulting density matrix invariably diagonalizes in the interaction basis. Of course, this does \textit{not} imply that the density matrix corresponding to the free field, in the absence of interactions, must also be diagonal. However, since interaction terms must be present (as these nonlinearities considered here are purely gravitational in nature), the density matrix of the full system must definitely become diagonal. Of course, adding the interaction term would also change the diagonal term. However, this change can, \`a priori, be assumed to be small since they are suppressed by factors of the interaction parameter and, in the first approximation, we calculated the entropy of the free-field density matrix by choosing to ignore the off-diagonal terms as part of our reduction (or, coarse-graining) procedure. Remarkably, we showed that the result was consistent with the classical calculations done earlier \cite{Tom1, Tom2} for the squeezing entropy. Thus, the role of nonlinearities in calculating this entropy, corresponding to the squeezed vacuum, is that of justifying our coarse-graining of ignoring the off-diagonal terms. In the following section, we will compute the entanglement entropy which the non-Gaussianities directly generate and show that our first-pass assumption that their effect, suppressed due the coupling parameter, is smaller than the squeezing entropy is not correct. Quite surprisingly, we shall find that the entanglement entropy, due to the same non-Gaussian term considered here, is larger than the squeezing entropy.

\section{Enhanced Entanglement Entropy due to Nonlinearities}\label{Ent_Entropy}

\subsection{Setup}
Having set up our interaction terms, let us discuss how one can calculate the entanglement entropy of the cosmological perturbations due to the effects of these coupling terms. To calculate the entanglement entropy, we shall follow the prescription of \cite{Bala}, and generalize their results for flat spacetime to inflationary backgrounds. 

Given our breakup of the Hilbert space \eqref{Hilbert_Space} $\mathcal{H} = \mathcal{H}_A \otimes \mathcal{H}_B$ into system and environment modes, our Hamiltonian can be expressed as 
\begin{eqnarray}
	H = H_A\otimes \mathbb{I} + \mathbb{I}\otimes H_B + \lambda H_{\rm int}\,,
\end{eqnarray}
where $H_{A,B}$ denote the free part of the Hamiltonian and $\lambda$ is a time-dependent constant. The ground state of the free theory, \textit{neglecting the interactions}, is denoted by $|0,0\rangle = |0\rangle \otimes |0\rangle$, and one can write the interacting vacuum of the entangled system as
\begin{eqnarray}
	|\Omega\rangle = &&|0,0\rangle + \sum_{n\neq 0} A_n |n,0\rangle + \sum_{n\neq 0} B_N |0,N\rangle\nn\\
	&&\;\;\;\;\;\;\;\;\;\; +\sum_{n,N\neq 0} C_{n,N} |n,N\rangle\, ,
\end{eqnarray}
where $|n\rangle$ denotes an n-particle state of the system (in fact, a product state over all super-Hubble $k$ modes), and $|N\rangle$ is the corresponding state for the bath.

Following the analyses of \cite{Bala}, one finds that the leading order contribution to the entanglement entropy for such a system can be written as 
\begin{eqnarray}\label{EE_formula}
	S_{\rm ent} = - \lambda^2 \log\(\lambda^2\) \sum_{n,N\neq 0} |\tilde{C}_{n,N}|^2\,,\nn\\
\end{eqnarray}
where we can express the matrix element $C_{n,N}$ in terms of standard perturbation theory as
\begin{eqnarray}\label{Matrix_element}
  \tilde{C}_{n,N} =  \dfrac{\langle n, N| H_{\rm int} |0,0\rangle}{(E_0 + \tilde{E}_0 - E_n - \tilde{E}_N)}\,.
\end{eqnarray}
For future convenience, we shall define the quantity $\tilde{C}_{n,N} = C_{n,N}/(E_0 + \tilde{E}_0 - E_n - \tilde{E}_N)$. Note that the crucial assumption which has been made above is that of time-independent perturbation theory, as was used in \cite{Bala} to calculate the matrix element $\tilde{C}_{n,N}$. Other than the explicit form for this matrix element, the formula for the entanglement entropy in \eqref{EE_formula} is completely general and applies to our case.  Of course, in order to calculate the entanglement entropy, we need to reinstate factors of the coupling parameter $\lambda (\eta) =\sqrt{\epsilon_H}/\(2\sqrt{2}\, a(\eta)\, \Mpl\)$. Since our interaction parameter, as well as the squeezed vacuum for the system modes, are both time-dependent, as we shall see later on, we should technically use time-dependent perturbation theory to calculate our matrix element. However, as shall be explicitly demonstrated in Appendix B, the leading order result remains unaltered from using the simple formula given above for the time-independent case. Therefore, in the following sub-sections, we shall continue to use the expressions \eqref{Matrix_element}, together with \eqref{EE_formula}, although the justification for that shall appear in Appendix B. 

Another complication is that in order to apply the above formula naively, it seems that we need the energy corresponding to the ground and excited states, both for the Minkowski and the squeezed vacuum, considered above. The trouble is that there is no well-defined notion for the energy of the squeezed vacuum. However, note that what we really need in the above formula is the energy difference between the first excited state and the corresponding vacuum, for both the Minkowski and the squeezed vacua. This is the same for both the system and the bath modes and is given by $\omega_k:= k$ for (nearly) massless scalar excitations. Thus, we need to replace $\(E_0 + \tilde{E}_0 - E_n - \tilde{E}_N\) = \(p_n + p_N\)$, the latter still being a well-defined quantity. This fact also plays a key role when we recall that the actual calculation which we should perform is that for time-dependent perturbation theory, as has been carried out in Appendix B, and not just for the approximation of the time-independent case as shall be treated in the following sections.

Before going on to calculate this matrix element, and the corresponding entanglement entropy for our cosmological system, let us review the flat space calculation first through an explicit example.

\subsection{Calculation for flat space}

Considering a cubic interaction term, one can write the action for a massive scalar field as
\begin{eqnarray}
	S = \int\d^4x \(-\frac{1}{2}\(\partial_\mu \varphi\)^2 - \frac{1}{2}m^2\varphi^2 - \frac{\lambda}{3!} \varphi^3\)\,. 
\end{eqnarray}
For a flat $(3+1)$-dimensional spacetime, the field can be decomposed in terms of the usual ladder operators as 
\begin{eqnarray}
\varphi(x) = \int  \frac{\d^3k}{ (2\pi)^3\,\sqrt{2 \omega_k}} \(a_\textbf{k} e^{-i\; \textbf{k}.\textbf{x}} + a^\dagger_\textbf{k} e^{i\; \textbf{k}.\textbf{x}}\)\,,
\end{eqnarray}
where $\omega_k = \sqrt{m^2+k^2}$. Here, instead of putting the fields in a box as in \cite{Bala}, we choose to work with continuous field variables, as would be more appropriate for cosmological perturbations later on. However, we still have a scale $\mu$ which separates our system from the environment, using the same convention as in \cite{Bala}. In other words, we are interested in calculating the entanglement entropy between the modes with momenta $k$ above and below $\mu$. In this case, the only nontrivial contribution to the matrix element would be from an excited state of a $3$-particle one which can be written as
\begin{eqnarray}
\left|p_1 p_2 p_3\right\rangle = a^\dagger_{\textbf{p}_1} a^\dagger_{\textbf{p}_2} a^\dagger_{\textbf{p}_3} \left|0\right\rangle\,.
\end{eqnarray}
Recalling that the interaction Hamiltonian is $\(\lambda/3!\) \varphi^3$, $\lambda$ having dimension of mass, the required matrix element \eqref{Matrix_element} can be written as
\begin{widetext}
	\begin{eqnarray}\label{Flat_Space}
C_{n,N}^{\rm flat} =	&&\int \d^3 x \left\langle p_1 p_2 p_3 \right| \[\int \frac{\d^3k}{(2\pi)^3\sqrt{2 \omega_k}} \(a_\textbf{k} e^{-i\; \textbf{k}.\textbf{x}} + a^\dagger_\textbf{k} e^{i\; \textbf{k}.\textbf{x}}\)\]^3 \left| 0 \right\rangle\nn\\
	= &&\int\d^3 x \left\langle p_1 p_2 p_3 \right| \[ \int\d^3k_1 \int\d^3k_2\int\d^3k_3\;\frac{1}{\(2\pi\)^9\sqrt{\omega_{k_1}  \omega_{k_2} \omega_{k_3}}} \(a^\dagger_{\textbf{k}_1} e^{i\; \textbf{k}_1.\textbf{x}}\)  \(a^\dagger_{\textbf{k}_2} e^{i\; \textbf{k}_2.\textbf{x}}\)  \(a^\dagger_{\textbf{k}_3} e^{i\; \textbf{k}_3.\textbf{x}}\)\] \left| 0 \right\rangle\nonumber\\
	= &&\frac{1}{2^{3/2}}\;\int\d^3 x  \[\frac{1}{\sqrt{\omega_{p_1}  \omega_{p_2} \omega_{p_3}}} e^{i\; \(\textbf{p}_1+\textbf{p}_2+\textbf{p}_3\).\textbf{x}} \]\nonumber\\
	= && \frac{1}{2^{3/2}}\;\frac{\(2\pi\)^3}{\sqrt{\omega_{p_1}  \omega_{p_2} \omega_{p_3}}}\; \delta^3\(\textbf{p}_1 + \textbf{p}_2 + \textbf{p}_3\)\,.
	\end{eqnarray}
\end{widetext}
In the second line above, we only keep the creation operators as required, whereas in the third line we have used the orthonormality property of the inner product to eliminate the integrals over $({\bf k}_1, {\bf k}_2 , {\bf k}_3)$. In the final step, we used the integration over the spatial coordinate, and the remaining delta function implies that at least one of the spatial momenta must be above, and at least one below, the scale demarcating the system and the environment.

The entanglement entropy for this system can be then evaluated by plugging in the above expression into \eqref{EE_formula}
\begin{eqnarray}
s_{\rm ent}^{\rm flat} &=&  -\lambda^2 \log\(\lambda^2\)\; \frac{1}{2^3\, (2\pi)^6} \times \\
&& \;\;\;\;\;\int_{\{p\}_\mu} \prod\, \d^3p_i\; \frac{\delta_{\textbf{p}_1 + \textbf{p}_2 + \textbf{p}_3}}{\omega_{p_1} \omega_{p_2} \omega_{p_3} \(\omega_{p_1} + \omega_{p_2} + \omega_{p_3}\)^2}\nn
\end{eqnarray}
where the integrals are over a set of momenta such that there can only be two configurations of interest -- either one of $\(p_1, p_2, p_3\)$ is greater than $\mu$  while the rest are below $\mu$, or two of them are above while one is below $\mu$. We have also divided the total entanglement entropy by the (infinite) volume to express it as an entanglement entropy density ($\equiv S_{\rm ent}^{\rm flat}/{\rm Vol}$).

\subsection{Vacuum \& Interaction Hamiltonian}
Let us first outline the differences we anticipate between the flat space calculation above and our case for cosmological perturbations. Firstly, the interaction parameter $\lambda = \lambda(\eta)$ will now be time-dependent. Secondly, the vacuum for the system modes is now given by the squeezed vacuum, and the mode functions corresponding to the vacuum in curved spacetime will have a different form of their momentum dependence. Since the vacuum of the super-Hubble modes will now be the squeezed vacuum, there now are contributions of terms with both creation and annihilation operators in our case. As mentioned earlier, these two reasons are responsible for making the system time-dependent. However, once again, we emphasize that our result, as shall be derived in the next sub-section, is valid up to the leading order term even though we use time-independent perturbation theory to derive it. Explicit proof that this is the case can be found in Appendix B. Finally, a major conceptual difference arises from the fact that the scale separating our system from the bath is given by the (comoving) Hubble scale which is time-dependent since we are working with comoving coordinates (and, in addition, by itself has a weak time-dependence of its physical value during inflation), and is not some arbitrary, tunable parameter $\mu$ as in the flat space case. With this in mind, let us begin by factoring the Hamiltonian for the overall system as 
\begin{eqnarray}
H = H_{\rm sys} + H_{\rm bath} + H_{\rm int}\,,
\end{eqnarray}
where the $H_{\rm sys}$ and $H_{\rm bath}$ is the quadratic Hamiltonian, for the super and sub-Hubble modes respectively, as given in \eqref{H_2}. Next, we write down the vacuum modes for the unperturbed systems, ignoring nonlinearities, as
\begin{eqnarray}\label{Curved_vacuum}
	|0, 0\rangle = |0\rangle_{k>aH} \otimes |SQ(\eta)\rangle_{k<aH}\,.
\end{eqnarray}
The $|0,0\rangle$ is the vacuum state for both the system as well as the bath modes. For the super-Hubble modes, the vacuum is given by the squeezed state as given in \eqref{Squeezed}. On the other hand, we have the usual Minkowski vacuum for the sub-Hubble modes, denoted by $|0\rangle$.

The explicit form of the interaction Hamiltonian naturally depends on the choice of the interaction term we choose between the perturbation modes. As mentioned earlier, for this paper, we shall restrict ourselves to only cubic perturbation terms which arise naturally from gravitational nonlinearities in any model of inflation, as captured by our interaction Lagrangian given in \eqref{CubicLag}. We emphasize once again that considering more complicated interactions or more fields can lead in a different term dominating $H_{\rm int}$, which would end up producing enhanced amounts of entanglement entropy. In this precise sense, we give a lower bound on the amount of entropy production coming from scalar modes during inflation.

For our dominant interaction term of the form 
\begin{eqnarray}
	\Mpl^2 \int \d t\,\d^3x\; a\, \epsilon_H^2\, \zeta \(\partial\zeta\)^2\,,
\end{eqnarray}
we can write down the interaction Hamiltonian by converting the $\zeta$ field to our canonical field $\chi$, and then expanding in terms of the creation and annihilation operators in momentum space. We find the following expression \cite{Gong:2019yyz}:
\begin{widetext}
	\begin{eqnarray}\label{Hint}
	\lambda(\eta)	H_{\rm int} &=& \lambda(\eta) \int_{\Delta}  \left[ \sqrt{\frac{k_2 k_3}{k_1}} \left(c^\dagger_{-\textbf{k}_1} c^\dagger_{-\textbf{k}_2} c^\dagger_{-\textbf{k}_3} + c_{\textbf{k}_1} c^\dagger_{-\textbf{k}_2} c^\dagger_{-\textbf{k}_3} +  \ldots \right) + \sqrt{\frac{k_2 k_1}{k_3}} \left(c^\dagger_{-\textbf{k}_1} c^\dagger_{-\textbf{k}_2} c^\dagger_{-\textbf{k}_3} +\ldots \right)\right. \nonumber\\
	 && \hspace{2cm} \left.+ \sqrt{\frac{k_1 k_3}{k_2}} \left(c^\dagger_{-\textbf{k}_1} c^\dagger_{-\textbf{k}_2} c^\dagger_{-\textbf{k}_3} + \ldots\right) \right]\,.
	\end{eqnarray}
\end{widetext}
where all the terms in the parentheses (\ldots) are the same and include all possible (momentum-conserving) combinations of the ladder operators. We have also defined $\int_\Delta := \int \frac{\d^3 k_1}{(2\pi)^3}\,\frac{\d^3 k_2}{(2\pi)^3}\,\frac{\d^3 k_3}{(2\pi)^3}\, \(2\pi\)^3 \delta^3(\textbf{k}_1+\textbf{k}_2+\textbf{k}_3)$. The difference in the momenta dependence of our choice of $H_{\rm int}$ from, say, one with time-derivatives such as $\mathcal{L}_3 \sim \zeta (\zeta')^2$, would be that some of the terms in the expression above would come with a minus sign since, in that case, the interaction term couples the field with its conjugate momentum \cite{Shandera:2017qkg}. The prefactor is given by (keeping in mind that we go from cosmic time to conformal time)
\begin{eqnarray}\label{Int_para}
	\lambda(\eta) = \frac{\sqrt{\epsilon_H}}{2\sqrt{2}\, a \Mpl}\,,
\end{eqnarray}
where, as anticipated, we get a time-dependent interaction parameter. We now have all the ingredients -- the vacuum state and the interaction Hamiltonian -- to calculate the matrix element given in \eqref{Matrix_element}.

\subsection{Matrix element}

Let us revisit our calculation of the matrix element for the cubic Lagrangian in Minkowski space. The crucial difference between that calculation and the one for inflation would be that instead of only keeping the term which solely involves creation operators from the interacting Hamiltonian, we shall also have to consider terms of the form $c_{\textbf{k}_1} c^\dagger_{-\textbf{k}_2} c^\dagger_{-\textbf{k}_3}$ and $c_{\textbf{k}_1} c_{\textbf{k}_2} c^\dagger_{-\textbf{k}_3}$. This is easy to understand since for the case of flat spacetime, the only nonzero contribution for the matrix element between the Minkowski vacuum and an excited state (with, say, three particles for a cubic interaction) can come if we sandwich a term consisting of three creation operators in between. If there exists any annihilation operator, it would simply annihilate the vacuum, resulting in zero. On the other hand, for inflation, we have a tensor product of the Minkowski vacuum for the sub-Hubble modes and the squeezed vacuum for thr super-Hubble ones \eqref{Curved_vacuum}. In this case, the ladder operator(s) corresponding to the sub-Hubble modes must be creation ones $c^\dagger_{-\textbf{k}}$ whereas the one(s) corresponding to the super-Hubble modes can be either $c^\dagger_{-\textbf{k}}$ or $c_{\textbf{k}}$. This is so because an annihilation operator $c_{\textbf{k}}$ does \textit{not} annihilate the squeezed vacuum $\left|SQ(k,\eta)\right\rangle$. One can see this explicitly from the form of the two-mode squeezed vacuum, as given in \eqref{2modeSqueezed}.

Having said this, let us list all the possible choices of interaction terms which can appear in the matrix elements:
\begin{itemize}
	\item Terms of the form $c^\dagger_{-\textbf{k}} c^\dagger_{-\textbf{k}} c^\dagger_{-\textbf{k}}$: There can be either two system (super-Hubble) modes and one bath (sub-Hubble) mode or vice-versa.
	\item Terms of the form $c_{\textbf{k}} c^\dagger_{-\textbf{k}} c^\dagger_{-\textbf{k}}$:  There can be either two system modes and one bath  mode or vice-versa. However, the annihilation operator must always correspond to the super-Hubble mode.
	\item Terms of the form $c_{\textbf{k}} c_{\textbf{k}} c^\dagger_{-\textbf{k}}$: There must be two system modes, corresponding to the two annihilation operators, and can, therefore, only be one bath mode.
    \item The terms proportional to $c_{\textbf{k}} c_{\textbf{k}} c_{\textbf{k}}$ necessarily yield zero for the matrix element since the annihilation operator corresponding to any of the bath modes annihilates the Minkowski vacuum.
\end{itemize}

Let us consider the first case in detail in the following calculation while we leave the details of the other terms for the Appendix A. Therefore, the term of interest for us from the $H_{\rm int}$ \eqref{Hint}, for calculating \eqref{Matrix_element}, is the following:
\begin{eqnarray}
	\left(\sqrt{\frac{k_2 k_3}{k_1}} + \sqrt{\frac{k_1 k_3}{k_2}} +\sqrt{\frac{k_1 k_2}{k_3}}\right) c^\dagger_{-\textbf{k}_1} c^\dagger_{-\textbf{k}_2} c^\dagger_{-\textbf{k}_3} \subset H_{\rm int}\,.\nn
\end{eqnarray}
Next, we need to find the explicit action of a creation operator on the squeezed vacuum. Using the definition of the two-mode squeezed state from \eqref{2modeSqueezed}, we can formally express the action of a creation operator on it as 
\begin{eqnarray}
	c^\dagger_{-\textbf{p}} \left|SQ\(k,\eta\) \right\rangle\,.
\end{eqnarray} 
Schematically, it implies that we are considering an excited state with a particle of energy $p$ over our squeezed vacuum.
A similar iteration would create higher order excited states over the squeezed vacuum. However, recall that for a cubic interaction  term, the only non-zero contribution to the matrix element comes from having the first excited state over both the squeezed and the Minkowski vacuum.  Also, since we are only considering cubic interactions, there can be only two choices --- either one of the modes is in the system and two are in the bath or two of them are in the system while one is in the bath. However, it will be clear from the following that the dominant contribution to the entanglement entropy comes from having two of the modes in the bath and one in the system. This is not at all surprising keeping in mind that the decoherence rate is also dominated by having two short-wavelength modes and one long-wavelength one.  

Let us consider the former option first, \textit{i.e.} $p_1, p_2 > a H $ while $p_3< a H$. The appropriate excited state to consider is of the form
	\begin{eqnarray}
	\left|n, N\right\rangle = \left| 1_{\textbf{-p}_{1}} \, 1_{\textbf{-p}_{2}} \right\rangle \otimes \;c^\dagger_{-\textbf{p}_3} \left|SQ\(k,\eta\) \right\rangle\,.
	\end{eqnarray}  
The only other novelty for our calculation is the effect of the squeezed vacuum on the inner product. Recall the standard result
\begin{widetext}
\begin{eqnarray}
	\left\langle SQ(k,\eta)\right| c_{\bf p} c_{\bf -q}^\dagger \left|SQ(k,\eta)\right\rangle &=& \left[ \left\langle SQ(k,\eta)\right|\left|S(k,\eta)Q\right\rangle  + \left\langle SQ(k,\eta)\right| N_{\bf p} \left|SQ(k,\eta)\right\rangle\right] \delta^3({\bf p} + {\bf q})\nn\\
	&=& \(1+\sinh^2 r_p\) \delta^3({\bf p} + {\bf q})\,,
\end{eqnarray}
where we have written things schematically to avoid clutter. To explicitly see how this result comes about, one should write down the unitary transformation of the creation and annihilation operator under the squeezing operator, \textit{i.e.} $S^\dagger c S$ and $S^\dagger c^\dagger S$ as linear combinations of $c, c^\dagger$, dropping all momenta indices. Also, note that $S^\dagger =S^{-1}$. See the Appendix A for more details.  The rest of the calculation follows exactly that of flat space, and it is easy to evaluate the matrix element as 
\begin{eqnarray}\label{C_sq}
	{}^{(c^\dagger c^\dagger c^\dagger)}C_{n,N}^{\rm sq} = 	(2\pi)^3 \(1+\sinh^2 r_{p_3}\)\left(\sqrt{\frac{p_2 p_3}{p_1}} + \sqrt{\frac{p_1 p_3}{p_2}} +\sqrt{\frac{p_1 p_2}{p_3}}\right) \; \delta^3(\textbf{p}_1 + \textbf{p}_2 + \textbf{p}_3)\,.
\end{eqnarray}
\end{widetext}
It is clear that for our choice of $p_1, p_2 \in \text{bath}$ while $p_3 \in \text{system}$, the dominant term in the above comes from the third term $\(C_{n,N} \propto \sqrt{\frac{p_1 p_2}{p_3}}\)$. It is also evident from the above calculation that if we had two modes in the system and one in the bath, then the dominant term in the matrix element would have the form 
\begin{widetext}
\begin{eqnarray}\label{C_fold}
{}^{(c^\dagger c^\dagger c^\dagger)}C_{n,N}^{\rm fold} &=& 	(2\pi)^3 \(1+\sinh^2 r_{p_2}\) \(1+\sinh^2 r_{p_3}\)\left(\sqrt{\frac{p_2 p_3}{p_1}} + \sqrt{\frac{p_1 p_3}{p_2}} +\sqrt{\frac{p_1 p_2}{p_3}}\right) \; \delta^3(\textbf{p}_1 + \textbf{p}_2 + \textbf{p}_3)\nn\\
&& \approx (2\pi)^3 \(1+\sinh^2 r_{p_2}\) \(1+\sinh^2 r_{p_3}\)\left( \sqrt{\frac{p_1 p_3}{p_2}} +\sqrt{\frac{p_1 p_2}{p_3}}\right) \; \delta^3(\textbf{p}_1 + \textbf{p}_2 + \textbf{p}_3)\,\,
\end{eqnarray}
\end{widetext}
where we have chosen $p_1 > aH$ and $p_2, p_3 < aH$. Already at this stage we can see that the entanglement entropy for cosmological perturbations, during inflation, peaks in the ``squeezed'' limit $p_3 \ll p_1 \approx p_2$, given the momentum structure of the matrix element, for ${}^{(c^\dagger c^\dagger c^\dagger)}C_{n,N}^{\rm sq}$ whereas it gets its maximum contribution in the ``folded'' limit $p_3 + p_2 \approx p_1$ for the other case ${}^{(c^\dagger c^\dagger c^\dagger)}C_{n,N}^{\rm fold}$.

\subsection{Entanglement entropy}

Let us recall the formula for the leading order term in the entanglement entropy 
\begin{eqnarray}
	S_{\rm ent}&=& -\lambda^2 \ln(\lambda^2)\;\; \sum_{n,N\neq 0} \dfrac{|C_{n,N}|^2}{\(p_n +p_N\)^2}\,,\nn\\
\end{eqnarray}
where a sum is implied on both types of $C_{n,N}$ calculated in \eqref{C_sq} and \eqref{C_fold}. Note our slight difference in convention of defining the matrix element $C_{n,N}$ with that of \cite{Bala} (our $\tilde{C}_{nN}$ in \eqref{Matrix_element} is equivalent to their $C_{nN}$). Also, we have replaced the explicit expressions for the energy eigenvalues in the original flat space formula by the comoving momenta corresponding to the energy difference. As mentioned, the energy difference between an excited state and the vacuum -- both Minkowski \& squeezed -- is still a well-defined quantity. This observation remains crucial in the calculation carried out in Appendix B for the time-dependent case, and not just for the approximation used here.

Note that the sum over $(n,N)$ translates into integrals over all the momentum modes in the formula \eqref{EE_formula}. Recall that there was a similar integral over all momentum modes also in the expression of the entropy arising from the squeezing part of the quadratic Hamiltonian, as shown in \eqref{Squeezed_entropy}. However, unlike in that case, we would have the integrals over all momentum conserving configurations involving $({\bf p_1}, {\bf p_2}, {\bf p_3})$ and not over individual modes as is expected for an entanglement entropy coming from cubic interactions. Keeping this is mind, the entanglement entropy (per unit comoving volume) is given by
\begin{widetext}
\begin{eqnarray}\label{S_final}
{}^{(c^\dagger c^\dagger c^\dagger)}s_{\rm ent} &=& -(2\pi)^3\lambda^2 \ln(\lambda^2)\;\; \int_H^{a H}\dfrac{\d^3p_3}{(2\pi)^3} \int_{aH}^{a\Mpl}\dfrac{\d^3p_2}{(2\pi)^3} \int_{a H}^{a\Mpl} \dfrac{\d^3p_1}{(2\pi)^3}\;\delta^3(\textbf{p}_1 + \textbf{p}_2 + \textbf{p}_3) \(\frac{p_1 p_2}{p_3}\) \; \frac{\(1+\sinh^2 r_{p_3}\)^2}{\(p_1 + p_2 + p_3\)^2} \nn  \\
&&-(2\pi)^3\lambda^2 \ln(\lambda^2)\;\; \int_H^{a H}\dfrac{\d^3p_3}{(2\pi)^3} \int_{H}^{aH}\dfrac{\d^3p_2}{(2\pi)^3} \int_{a H}^{a\Mpl} \dfrac{\d^3p_1}{(2\pi)^3}\;\delta^3(\textbf{p}_1 + \textbf{p}_2 + \textbf{p}_3) \times \nn\\
&& \hspace{2cm}\[\left(\sqrt{\frac{p_1 p_3}{p_2}} +\sqrt{\frac{p_1 p_2}{p_3}}\right)^2\; \frac{\(1+\sinh^2 r_{p_2}\)^2\,\(1+\sinh^2 r_{p_3}\)^2}{\(p_1 + p_2 + p_3\)^2}\] =: I_1 + I_2\,,
\end{eqnarray}
\end{widetext}
where we have only kept the dominant terms from the matrix elements \eqref{C_sq} and \eqref{C_fold}. It is important to discuss the limits of the above integral first: We have introduced $\Mpl$ as the natural \textit{physical} UV cutoff and the comoving wavenumber at the beginning of inflation as the infrared cutoff. We set $a_i=1$ for the scale factor at the beginning of inflation (and therefore, in our convention, $a$ is always $>1$). We also assume that the Hubble parameter, $H$, remains constant during inflation. Furthermore, the UV cutoff for the comoving momenta is given by $a\Mpl$ which signifies the fact that the integration of the environment is over a fixed number of bath modes, even though we are considering an accelerating background. This is so because although the environment is continuously depleted by modes getting redshifted into the system, there is also a constant supply of modes from the UV into the bath\footnote{This mode creation is a source of non-unitarity which is one of the arguments for the TCC \cite{TCC,TCC2}.}. However, the system has an increasing phase space of modes as more and more modes become super-Hubble as time goes on, and given our infrared cutoff which states that there were no comoving modes which were super-Hubble before inflation started. Naturally, we have to assume that inflation starts at a finite time in the past which reinforces the need of having an UV cutoff for the perturbation modes.

Let us now estimate the integrals $I_1$ and $I_2$ given in \eqref{S_final}. For $I_1$, when we have two bath modes and one system mode, the integrand would naturally have its largest contribution coming from the squeezed limit, as shown below:
\begin{widetext}
\begin{eqnarray}\label{I_1}
I_1 &=& -(2\pi)^3\lambda^2 \ln(\lambda^2)\;\; \int_H^{a H}\dfrac{\d^3p_3}{(2\pi)^3} \int_{aH}^{a\Mpl}\dfrac{\d^3p_2}{(2\pi)^3} \int_{a H}^{a\Mpl} \dfrac{\d^3p_1}{(2\pi)^3}\;\delta^3(\textbf{p}_1 + \textbf{p}_2 + \textbf{p}_3) \(\frac{p_1 p_2}{p_3}\) \; \frac{\(1+\sinh^2 r_{p_3}\)^2}{\(p_1 + p_2 + p_3\)^2}\nn\\
&=& -\lambda^2 \ln(\lambda^2) \int_H^{a H}\dfrac{\d^3p_3}{(2\pi)^3} \int_{aH}^{a\Mpl}\dfrac{\d^3p_2}{(2\pi)^3} \(\frac{p_2 \sqrt{p_2^2 + p_3^2 +2 p_2 p_3 \cos\Theta}}{p_3}\)\, \dfrac{\(1+\sinh^2 r_{p_3}\)^2}{\(\sqrt{p_2^2 + p_3^2 +2 p_2 p_3 \cos\Theta} + p_2 + p_3\)^2}\nn\\
&\approx& -\lambda^2 \ln(\lambda^2)\int_{aH}^{a\Mpl} \dfrac{\d^3p_2}{(2\pi)^3} \int_H^{a H}\dfrac{\d^3p_3}{(2\pi)^3}\; \dfrac{\(a H\)^4}{2^4\,p_3^5}\nn\\
&\sim& \dfrac{\epsilon_H}{3\, \(2\pi\)^4\, 2^6\, a^2 \Mpl^2} \, \(aH\)^4 \[\(a\Mpl\)^3 - \(aH\)^3\] \; \[\frac{1}{H^2} - \frac{1}{\(aH\)^2}\] \times \ln\(\lambda^2\) \;\lesssim\; \epsilon_H\; H^2 \; \Mpl \;a^5\; \ln(\lambda^2)\,.
\end{eqnarray}	
\end{widetext}
In the second line, we have killed the $p_1$ integral using the delta function, introducing the angle $\Theta$ between ${\bf p}_2$ and ${p}_3$. In the next line, we introduce the crucial approximation that the integrand peaks in the limit $\Theta \rightarrow \pi/2$ and $p_2 \gg p_3$, \textit{i.e.} the squeezed limit. This would help us in getting an upper bound on the entanglement entropy corresponding to the $I_1$ term. We have also used the expression for the squeezing parameter from \eqref{r_k} and used the approximation that $1+\sinh r_k \approx \sinh r_k$, for large squeezing, in this step. It is then easy to see that the integration over the bath modes is dominated by the upper limit (the UV cutoff scale), while the integral over the system mode $p_3$ is dominated by the lowest value of $p_3$, i.e. by the infrared (IR) cutoff scale. We have only kept the leading terms in the integrals in the same spirit to arrive at our lower estimate for the entropy density, ignoring numerical factors. We note that a factor of  $a^3$ should be divided from the final result in order to account for the entanglement entropy density (total entropy per unit \textit{physical} volume). We are then left with a factor of $\(a/a_i\)^2$ (recall, we have set $a_i=1$) and this reflects the fact that the phase space of the system modes is growing, and the contribution to the ${p}_3$ integral is dominated by the IR cutoff. Collecting everything, the estimate\footnote{To remind the readers, this is a lower bound on the amount of entanglement entropy produced in any model of inflation since we are only considering cubic interactions of density perturbations alone, which come from minimally coupling a scalar field to GR. There are necessarily other sources such as those due to non-Gaussian terms for tensor perturbations.} of the entanglement entropy per unit physical volume coming from $I_1$ is given by
\begin{eqnarray}\label{Estimate1}
 s^{I_1}_{\rm ent} \lesssim \epsilon_H\; H^2 \; \Mpl \;a^2\; \ln(\lambda^2)\,,	
\end{eqnarray} 
where $a>1$ is such that the number of $e$-foldings of inflation is given by $N:= \ln a$ in our convention. 

Let us now first show that the contribution coming from $I_2$ to the entanglement entropy density would be subdominant to the above result. In this case of having two system and one bath mode, the largest contribution to the integrand would come from the folded limit $p_1 \approx p_2 + p_3$. Following the calculation as in the previous case, we can arrive at an upper bound for the estimate of this term in a similar way. However, note that once we eliminate the integral over the bath mode $p_1$ using the delta function, none of the system mode integrals which are left have any dependence on $\Mpl$. The other difference lies in the additional squeezing terms leading to an extra factor of the IR cutoff in the final result, namely, 
\begin{eqnarray}
 s^{I_2}_{\rm ent} \lesssim \epsilon_H\; H^5 \; \frac{1}{\Mpl^2} \;a^3\; \ln(\lambda^2)\,.
\end{eqnarray}
Once again, we have expressed this final result in terms of the entanglement entropy per unit physical volume and have only given a rough estimate of the upper bound. Thus, we find
\begin{eqnarray}
	f:= \frac{s^{I_1}_{\rm ent}}{s^{I_2}_{\rm ent}} = \frac{1}{a} \;\(\frac{\Mpl}{H}\)^3\,,
\end{eqnarray}
which means that $s^{I_2}_{\rm ent}$ shall always remain subdominant to $s^{I_1}_{\rm ent}$, provided $f>1 \Rightarrow N < 3\ln(\Mpl/H)$. In the next section, we shall show that combining the observed scalar power spectrum with the fact that the entanglement entropy of cosmological perturbations during inflation remain smaller than the thermal entropy produced during (p)reheating leads to this condition being always satisfied. Therefore, we can always ignore the entanglement entropy corresponding to having two system and one bath mode when compared to that of having two sub- and one super-Hubble mode. 

Note that the above estimates were calculated using the approximations of squeezed and folded shapes, in which the integrands reach their peak values. The full integrals do not lend themselves to having simple analytic forms and we have thus avoided writing them down explicitly. The effect of removing these approximations would result in some small numerical factors appearing in front of our estimates, as in \eqref{Estimate1}. However, recall that we have only shown here the result of the calculation of the entanglement entropy coming from the terms of the form $c^\dagger_{-\textbf{k}} c^\dagger_{-\textbf{k}} c^\dagger_{-\textbf{k}}$, arising from the interaction Hamiltonian in \eqref{Hint}. As mentioned earlier, there are other terms, proportional to $c_{\textbf{k}} c^\dagger_{-\textbf{k}} c^\dagger_{-\textbf{k}}$ and $c_{\textbf{k}} c_{\textbf{k}} c^\dagger_{-\textbf{k}}$, which also contribute to the entanglement entropy. As shown in the Appendix A, in the limit of large squeezing, $r_k \gg 1$, the contribution of all of these terms are either proportional to $s^{I_1}_{\rm ent}$ or to $s^{I_2}_{\rm ent}$. Naturally, we neglect the terms proportional to $s^{I_2}_{\rm ent}$ since they are sub-dominant. And the terms which are proportional to $s^{I_1}_{\rm ent}$ shall add to our estimate for the entanglement entropy density \eqref{Estimate1}. All of this is to say that in our order of magnitude estimate for the entanglement entropy density of cosmological perturbations during inflation, there should be some $\mathcal{O}(1)$ numerical factor appearing, namely
\begin{eqnarray} \label{final}
	s_{\rm ent} \sim \mathcal{O}(1)\; \ln(\lambda^2)\;\epsilon_H\; H^2 \; \Mpl \;a^2\,.
\end{eqnarray}
There are two sources which contribute to this $\mathcal{O}(1)$ number -- one from the additional terms, as shown in the Appendix A, and the other coming from the fact that we are estimating the integral by its upper bound. From now on, we shall drop this number as well as the logarithmic factor in our upcoming discussions.

Now that we have an estimate for the entanglement entropy due to the gravitational nonlinearities, let us compare this with the contribution coming from the squeezing part of the quadratic Hamiltonian, as in \eqref{Squeezed_entropy}. As mentioned earlier, for large $r_k \gg1$, the entropy density (per physical volume), coming from \eqref{Squeezed_entropy}, is given by \eqref{Entropy_Squeezed_estimate}
\begin{eqnarray}
	s_{\rm sq} &=& \frac{1}{a^3} \, \int_{H}^{aH} \d^3 k \ln\(\sinh^2 r_k\) \sim H^3\,,
\end{eqnarray}
where we have, once again, ignored some small numerical factors. 

Although $s_{\rm ent}$ corresponding to cubic interactions arising from gravitational nonlinearities is suppressed by a factor of $\epsilon_H$ (as it should be), it is still greater than $s_{\rm sq}$. One way to easily see this is to approximate the value of the observed scalar power spectrum as
\begin{eqnarray} \label{powersp}
 P_{\zeta} \sim \frac{1}{\epsilon_H}\,\(\dfrac{H}{\Mpl}\)^2 \sim 10^{-9}\,,
\end{eqnarray}
such that $\epsilon_H \sim 10^9\, \(H/\Mpl\)^2$. Let us define the ratio
\begin{eqnarray}\label{f2}
	t := \frac{s_{\rm ent}}{s_{\rm sq}} \sim \epsilon_H\; \(\frac{\Mpl}{H}\)\; a^2    \sim 10^9\; \(\frac{H}{\Mpl}\)\; e^{2N}\,.
\end{eqnarray}
As we shall see from the bounds on $N$ that we will derive in the next section, this quantity $t >1$ and thus the entanglement entropy from non-Gaussianities would be larger than that corresponding to the squeezed vacuum, provided inflation lasts a reasonable amount of time and is not fine-tuned to be extremely small. This is quite a remarkable result since this implies that the entanglement entropy due to (cubic) gravitational nonlinearities are larger than that due to the (squeezing part of the) quadratic action! 

\section{Upper bound on the duration of inflation}\label{TCC}

We have seen that the entanglement entropy density of cosmological perturbations produced by nonlinearities builds up during a period of inflation as 
\begin{eqnarray}
\frac{a}{a_i} \, = \, e^{N} \, ,
\end{eqnarray}
where $N$ is the number of e-foldings of inflation, and $a_i$ is the value of the scale factor at the beginning of inflation (which we had set equal to $1$ in the last section, for simplicity). In order to allow a graceful exit from inflation consistent with the second law of thermodynamics,  it is important to make sure that the entropy due to these interactions remain subdominant to the entropy in the thermal radiation state after inflation. This thermal entropy density is given by
\begin{eqnarray}
s_{\rm{th}} \, = \frac{4 \pi^2}{45} g^* T_R^3 \, ,
\end{eqnarray}
where $T_R$ is the initial temperature of the radiation bath, and $g^{*}$ is the number of spin degrees of freedom in the radiation bath. Assuming rapid thermalization after inflation, and nearly constant Hubble parameter during inflation, this yields
\begin{eqnarray}
s_{\rm{th}} \, \simeq \, \frac{4 \pi^2}{45} g^* H^{3/2} \Mpl^{3/2} \, .
\end{eqnarray}
Making use of the result (\ref{final}), the requirement 
\begin{eqnarray}
s_{\rm{th}} \, > \, s_{\rm ent}
\end{eqnarray}
yields the condition
\begin{eqnarray}
N \, < \, \frac{1}{4} \ln{\(\frac{\Mpl}{H}\)} + \frac{1}{2} \ln \epsilon_H^{-1}
\end{eqnarray}
(modulo numerical factors). The value of $\epsilon_H$ is given in terms of $H$ and $\Mpl$ via the equation (\ref{powersp}), invoking the observed value of  the amplitude of the power spectrum of cosmological perturbations. Inserting the resulting relation for $\epsilon_H$ yields
\begin{eqnarray}\label{bound}
N \, < \, \frac{5}{4} \ln{\(\frac{\Mpl}{H}\)} - \frac{9}{2} \ln 10 \, ,
\end{eqnarray}
which is very close the bound \cite{TCC2}
\begin{eqnarray}
N \, < \, \ln{\(\frac{\Mpl}{H}\)}
\end{eqnarray}
which results from the TCC \cite{TCC}. Note that this bound on the duration of the inflationary phase is the same as derived in \cite{Dvali1}, where it was argued that beyond that time the de Sitter phase cannot be given a well-defined classical background interpretation due to the buildup of entanglement\footnote{In a later paper \cite{Dvali2}, another (and much larger) time scale was introduced as the time scale beyond which the actual de Sitter background breaks down. It was then argued \cite{Dvali3} that low energy effective field theory remains valid up to that time.}.

We are thus led to speculate the the TCC may have a derivation based on entropy considerations and the second law of thermodynamics. It is already known that entropy considerations have also proven useful \cite{Shiu} to derive the de Sitter swampland conjecture \cite{swamp,Krishnan}, one of the various constraints on effective field theories to be consistent with string theory (see e.g. \cite{Vafa,Palti} for reviews).

Note that we have derived a lower bound on the entanglement entropy due to the minimal gravitational nonlinearities (ignoring those due to tensor perturbations). We might speculate that if we were to do a more detailed calculation, our entropy bound on $N$ might turn out to be in even closer agreement with the bound from the TCC. Note that the bound (\ref{bound}) can be relaxed if we consider $H$ to be decreasing substantially during inflation, or if the thermal history of the universe after inflation is non-standard. However, as shown in \cite{Kamali, Shi}, in these cases the TCC bound is also relaxed. Note, also, that if we take into account entanglement entropy due to modes which were already super-Hubble at the beginning of inflation, the bound can be strengthened, in the same way that the TCC bound is strengthened if we consider pre-inflation evolution \cite{Ed, Shi}. Finally, it has also been pointed out that deriving the TCC from different quantum gravity arguments can, by itself, lead to a refinement of it \cite{TCC_SDC} and can bring it closer to our bound.

Returning to the discussion at the end of the previous section concerning the ratio of the entropies produced by nonlinear entanglement effects on one hand, and by pure decoherence of the linear modes on the other, we see that if the duration of inflation saturates the above bound (\ref{bound}), then the entanglement entropy dominates by a factor of $(\Mpl / H)^{3/2}$, the result we promised to derive earlier. In other words, unless inflation lasts for a very short period of time, $s_{\rm ent}$ would always dominate over $s_{\rm sq}$.

Note that a related bound on the duration of inflation based on entanglement considerations was given in \cite{Bao}, where it was argued that, interpreting the current horizon entropy of the Universe as entanglement entropy, there is a number of e-foldings of inflation before which there is no entropy and we cannot talk about a de Sitter background.

\section{Conclusions and Discussion}\label{conclusion}

In this work, we have derived the entanglement entropy of inflationary scalar perturbations, corresponding to nonlinearities arising from gravity. Although entropy of cosmological perturbations is a rich subject by itself, what is novel to our work is that we calculate the \textit{entanglement} entropy to the leading order of cubic interactions, going beyond the calculation of entropy corresponding to the squeezing of the super-Hubble vacuum state. Remarkably, we show that this cubic (and higher order) interactions are essential even to calculate the entropy corresponding to the quadratic Hamiltonian. This is so because decoherence arising from these terms is what is responsible for reducing the pure density matrix to a mixed one, by suppressing the off-diagonal terms. These higher order interaction Hamiltonians themselves lead to mode-couplings such that there is an entanglement between the super- and sub-Hubble modes which is a direct manifestation of the quantum origin of these vacuum fluctuations\footnote{This property of the entanglement entropy corresponding to the interactions alone is something unique for models of the early-universe which explain macroscopic perturbations as originating from quantum vacuum fluctuations, unlike the entropy corresponding to the squeezing of the modes which can also be interpreted as some type of classical Shannon entropy.}.The entanglement entropy corresponding to these interactions is what we have calculated for the first time by treating the super-Hubble modes as our system and the sub-Hubble ones as a bath. 

Our result shows that the entanglement entropy density scales as $H^2 \Mpl \(a/a_i\)^2$, where $a_i$ is the scale factor at the beginning of inflation. In order to allow for a graceful exit from inflation consistent with the second law of thermodynamics, this entropy must be smaller than the thermal entropy after inflation. This leads to an upper bound on the duration of inflation which is very close to the bound obtained from the TCC. Interestingly, the nonlinearities produce the dominant contribution to the entropy of cosmological perturbations, surpassing the one for the squeezed vacuum, provided $\epsilon > \(H/\Mpl\) \, \(a_i/a\)^2$ and is \textit{not fine-tuned to be extremely small}. Using the upper bound derived on the duration of inflation, this translates into the statement that the entanglement entropy due to cubic interactions dominate over the one due to the (quadratic) squeezing term, provided inflation \textit{does not} last for a very short period of time. \`A priori, there is no reason to expect this to be the case and indeed one would intuitively guess that the squeezing entropy would dominate over the (cubic) entanglement entropy. As an aside, we rederived the squeezing entropy from the full quantum density matrix, using a suitable coarse-graining scheme, which match previous results \cite{Tom1,Tom2}, calculated using a stochastic classical field approximation, in the large squeezing (classical) limit.

As we have shown, the calculation of the entanglement entropy of cosmological perturbations simplifies when done in momentum space. It is easy to appreciate this properly if one compares our result with that for determining the full non-unitary evolution of the density matrix of the system modes as has been done, for instance, in \cite{Shandera:2017qkg} (see \cite{Gong:2019yyz} for the case of tensor modes). The time evolution of the reduced density matrix involves non-Hamiltonian terms, and might even contain  non-Markovian terms, which depend on the so-called Lindblad operator. If one were to try and calculate the solution of the time-dependent reduced density matrix and then evaluate the von Neumann entropy associated with it, the calculation would become much harder and rather intractable. In this paper, we give a complementary way of calculating the entanglement entropy without having to deal with the full dynamics since, as emphasized earlier, we only require to calculate certain matrix elements for our purposes. The fact that these two seemingly different methods yield the same result for the entanglement entropy has been shown in \cite{Agon:2014uxa} for any quantum field theory. In addition, going to momentum space makes it easy to impose a UV cutoff for the bath modes, as has been done in this case.

The natural next step for us would be to calculate the entanglement entropy corresponding to primordial gravitational waves. Once again, assuming the simplest model of inflation, nonlinearities would arise from gravitational interactions which would lead to decoherence and entropy production. Therefore, this calculation would also give an improved lower bound on the amount of entropy which must be produced in any model of inflation. Furthermore, the leading interactions between the tensor perturbations are \textit{not} slow-roll suppressed which typically lead them to decohere faster than their scalar counterpart \cite{Gong:2019yyz}. Anticipating along similar lines, we expect that the entanglement entropy of tensor modes would be somewhat enhanced, and this will be studied in future work. The cubic interactions coupling tensor and scalar modes also need to be taken into account which will result in enhancing both the entanglement entropy density of the scalar as well as the tensor perturbations.

Finally, we note that our analysis has been done in the context of inflationary cosmology, but the methods also apply to other early universe scenarios in which the primordial fluctuations are quantum in origin, in particular to the {\it matter bounce} and to the {\it Ekpyrotic} scenarios.

\vspace{5mm}
\section*{Acknowledgements}
\noindent RB thanks the  Pauli Center and the Institutes of Theoretical Physics and of Particle- and Astrophysics of the ETH for hospitality. The research at McGill is supported, in part, by funds from NSERC and from the Canada Research Chair program. SB is also supported in part by a McGill Space Institute fellowship and by a generous gift from John Greig. OA acknowledges the generous support of the McGill-UAE Graduate Studies Fellowships.
\\

\section*{Appendix A: Full Entanglement entropy}\label{App:A}

In the main body of the paper, we have shown in detail the derivation of the entanglement entropy due to the $c^\dagger_{-\textbf{k}} c^\dagger_{-\textbf{k}} c^\dagger_{-\textbf{k}}$ terms coming from the interaction Hamiltonian in \eqref{Hint}. However, as mentioned earlier, there are other terms which also contribute to the entropy. Let us first consider the terms of the form $c_{\textbf{k}} c^\dagger_{-\textbf{k}} c^\dagger_{-\textbf{k}}$ appearing in \eqref{Hint}:
\begin{widetext}
\begin{eqnarray}\label{A1}
\[c_{\textbf{p}_1} c^\dagger_{-\textbf{p}_2} c^\dagger_{-\textbf{p}_3} + c_{\textbf{p}_2} c^\dagger_{-\textbf{p}_1} c^\dagger_{-\textbf{p}_3} + c_{\textbf{p}_3} c^\dagger_{-\textbf{p}_2} c^\dagger_{-\textbf{p}_2}\] \(\sqrt{\dfrac{p_1 p_2}{p_3}} + \sqrt{\dfrac{p_1 p_3}{p_2}} + \sqrt{\dfrac{p_2 p_3}{p_1}}\) \,.
\end{eqnarray}
\end{widetext}
For terms such as these, we can have two possibilities as before -- two sub-Hubble modes and one super-Hubble mode or the other way around. Let us take the former case first. In this case, if $p_1, p_1 > aH$ and $p_3 < aH$, then the first term proportional to $\sqrt{\dfrac{p_1 p_2}{p_3}}$ would naturally be the dominant one. For this case, the only term which contributes would be the last one, proportional to $c_{\textbf{p}_3}$. This is a crucial argument, so let us emphasize it again -- the matrix element can be nonzero if there is no annihilation operator present in the inner product corresponding to sub-Hubble modes. The reason for this is the same as why there were no annihilation elements present in the inner product for the flat space calculation.

In this case, we need to calculate an inner product of the form
\begin{widetext}
\begin{eqnarray}\label{A3}
\left\langle SQ(k,\eta)\right| c_{\bf p} c_{\bf q} \left|SQ(k,\eta)\right\rangle &=& \left\langle0_\textbf{k}, 0_\textbf{-k}\right|S^\dagger_k\(r_k,\phi_k\) c_{\bf p} c_{\bf q} S_k\(r_k,\phi_k\) \left|0_\textbf{k}, 0_\textbf{-k}\right\rangle \nn\\
&=& - e^{i \phi_p}\, \cosh r_p\, \sinh r_p\, \delta^3({\bf p} + {\bf q})\,.
\end{eqnarray}
\end{widetext}
In deriving this, we have used the transformation of the annihilation operator under the unitary action of the squeezing operator, namely \cite{Holstein:1940zp} 
\begin{eqnarray}
	S^{-1} a S = a \cosh r  + a^\dagger e^{i\phi} \sinh r\,,
\end{eqnarray}
where we have dropped the momentum indices for simplicity. We have also used the fact that $S^\dagger = S^{-1}$.

The matrix element corresponding to this term would be given by
\begin{widetext}
\begin{eqnarray}
	{}^{\(cc^\dagger c^\dagger\)}C_{n,N}^{\rm sq} \sim - (2\pi)^3 \(e^{i \phi_{p_3}}\, \cosh r_{p_3}\, \sinh r_{p_3}\)\;\sqrt{\dfrac{p_1 p_2}{p_3}}\; \delta^3\(\textbf{p}_1 + \textbf{p}_2 + \textbf{p}_3\)\,.
\end{eqnarray}
\end{widetext}
Now let us recall that what enters in the formula of the entanglement entropy is not ${}^{\(cc^\dagger c^\dagger\)}C_{n,N}^{\rm sq}$ but rather its amplitude squared, \textit{i.e.}  $\left|{}^{\(cc^\dagger c^\dagger\)}C_{n,N}^{\rm sq}\right|^2$. In the limit of large squeezing, $\sinh r_{p_3} \approx \cosh r_{p_3} \gg 1$, and it is easy to see that the entanglement entropy corresponding to this term would be the same as that coming from $s^{I_1}_{\rm ent}$, as in \eqref{Estimate1}. 

Let us now return to our other possibility of having two super-Hubble modes $p_2, p_3 < aH$ and one sub-Hubble mode $p_1 > aH$. In this case, once again, the only nonzero contribution comes from the term proportional to $c_{{\bf p}_3}$ in \eqref{A1}. Of course now one of the creation operators, $c^\dagger_{{\bf p}_2}$, corresponds to a super-Hubble mode and thus we have an inner product of the form $\left\langle SQ(k,\eta)\right| c_{\bf p} c^\dagger_{-{\bf q}} \left|SQ(k,\eta)\right\rangle$ in addition to the one appearing in \eqref{A3}. Collecting these terms, the matrix element can easily be calculated to give
\begin{widetext}
\begin{eqnarray}
	{}^{\(cc^\dagger c^\dagger\)}C_{n,N}^{\rm fold} \sim - (2\pi)^3 \(e^{i \phi_{p_3}}\, \cosh r_{p_3}\, \sinh r_{p_3}\) \; \left(1+\sinh^2 r_{p_2}\right)\; \left( \sqrt{\frac{p_1 p_3}{p_2}} +\sqrt{\frac{p_1 p_2}{p_3}}\right)\; \delta^3\(\textbf{p}_1 + \textbf{p}_2 + \textbf{p}_3\)\,.
\end{eqnarray}
\end{widetext}
Once again, it is easy to see that in the limit $r_{p_3} \gg 1$, the contribution of this term to the entanglement entropy would be exactly the same as that of $s^{I_2}_{\rm ent}$. Thus, the contribution of this term would be subdominant, for the same reason as that of $s^{I_2}_{\rm ent}$.

Finally there remains one last type of terms which arise from the interaction Hamiltonian \eqref{Hint}, which are proportional to $c_{\bf k} c_{\bf k} c^\dagger_{-{\bf k}}$. These are the terms which go as
\begin{widetext}
\begin{eqnarray}\label{A2}
\[c_{\textbf{p}_1} c_{\textbf{p}_2} c^\dagger_{-\textbf{p}_3} + c_{\textbf{p}_1} c_{\textbf{p}_3} c^\dagger_{-\textbf{p}_2} + c_{\textbf{p}_3} c_{\textbf{p}_2} c^\dagger_{-\textbf{p}_1}\] \(\sqrt{\dfrac{p_1 p_2}{p_3}} + \sqrt{\dfrac{p_1 p_3}{p_2}} + \sqrt{\dfrac{p_2 p_3}{p_1}}\) \,.
\end{eqnarray}
\end{widetext}
For such terms, the only nonzero contribution appears when there are two super-Hubble and one sub-Hubble mode. In this case, there shall appear two factors of the inner product $\left\langle SQ(k,\eta)\right| c_{\bf p} c_{\bf q} \left|SQ(k,\eta)\right\rangle$ in the matrix element ${}^{\(c c c^\dagger\)}C_{n,N}^{\rm fold}$. It should be clear from the calculations above that the entanglement entropy corresponding to this term shall be the same as $s^{I_2}_{\rm ent}$ and shall, therefore, be sub-dominant. Once again, we have assumed the large squeezing limit to arrive at this conclusion.

\section*{Appendix B: Time-dependent perturbation theory}
We begin with the matrix element, for a time-dependent perturbation Hamiltonian, up to leading order:
\begin{eqnarray}\label{T-dep}
 \tilde{C}_{nN} = -i \int_{\eta_0}^{\eta} \d \eta' e^{i\omega \eta'}\, C_{nN}(\eta')\,,
\end{eqnarray}
where $C_{nN}(\eta) = \langle n, N| H_{\rm int}(\eta) |0,0\rangle$ and $\omega$is the energy difference between the states for which the matrix element is being calculated. Like before, one of the vacuum states, namely for the super-Hubble modes, is going to be the squeezed vacuum which is, by itself, also time-dependent. Furthermore, note that $\omega$ for our purposes is the energy difference between an one-particle state and the vacuum which is  well-defined and, just as before, is given by $\omega = p_1 + p_2 +p_3$.
 
Here, we shall carry out the explicit calculation for the ${}^{(c^\dagger c^\dagger c^\dagger)}s_{\rm ent}$ term for the case in which there are two sub-Hubble modes and one super-Hubble mode. In our notation, this should correspond to the $s_{\rm ent}^{I_1}$ result. From this calculation, it would be clear that using the time-dependent perturbation theory for the other terms would lead to the same result, up to the leading order term. Following \eqref{C_sq}, we find that (although there is no $I_1$ integral here so to speak, we keep this notation for this term to facilitate comparison with our earlier calculation):
\begin{widetext}
\begin{eqnarray}
s_{\rm ent}^{I_1}&\sim& (2\pi)^3\;\; \int_H^{a H}\dfrac{\d^3p_3}{(2\pi)^3} \int_{aH}^{a\Mpl}\dfrac{\d^3p_2}{(2\pi)^3} \int_{a H}^{a\Mpl} \dfrac{\d^3p_1}{(2\pi)^3}\;\delta^3(\textbf{p}_1 + \textbf{p}_2 + \textbf{p}_3) \(\frac{p_1 p_2}{p_3}\)\times\nn\\
&& \hspace{4cm}\left|\int_{\eta_0}^\eta\,\d\eta'\; e^{i\(p_1 + p_2 + p_3\) \eta'}\lambda \; \(1+\sinh^2 r_{p_3}\)\right|^2\,,
\end{eqnarray}
\end{widetext}
Let us slowly examine how we arrive at the above expression for the time-dependent case. Firstly, note that we have dropped the logarithmic term  since it would be small and this is consistent with our assumption earlier where we had also ignored this term, along with some numerical factors. The important observation is that both $\lambda$ and $r_{p_3}$ are time-dependent quantities and cannot be taken outside the time integral which appears in \eqref{T-dep} (the factor of `$-i$' does not make any difference since we consider the absolute value of the matrix element). This, and the fact, that there is no $\(p_1 + p_2 + p_3\)$ in the denominator is what distinguishes this expression from our simplified time-independent assumption earlier. Let us focus on the time-integral first and use \eqref{squeezing_paramters} and the fact that we have large squeezing, to get
\begin{eqnarray}
\int_{\eta_0}^\eta \d\eta' \, \dfrac{e^{i\(p_1 + p_2 + p_3\) \eta'}}{4 p_3^2\eta'}\, \(\dfrac{H \sqrt{\epsilon_H}}{2\sqrt{2}\Mpl}\)\,.
\end{eqnarray}
We can pull out everything which is time-dependent outside this integral, remembering to square everything. (We have also gotten rid of the $(-i)$ factor appearing in the definition of the matrix element since we only need its absolute value.) However, the crucial part is indeed the time-intergral which is now approximated as
\begin{eqnarray}
	\int_{\eta_0}^\eta \d\eta' \, \dfrac{e^{i\(p_1 + p_2 + p_3\) \eta'}}{\eta'}
	\approx \frac{1}{\eta}\; \frac{1}{\(p_1+p_2+p_3\)}\,. 
\end{eqnarray}
Let us go over this approximation slowly as this is the most important result for us in this calculation. By the Riemann-Lebesgue lemma, $\int_{\eta_0}^\eta \d\eta' f(\eta') e^{i\(p_1 + p_2 + p_3\) \eta'} = \mathcal{O}\(1/\(p_1 + p_2 + p_3\)\)$ when $\(p_1 + p_2 +p_3\) \rightarrow \infty$. This is true since, for us, $f(\eta') = 1/\eta'$, a $C^\infty$ function over $\(\eta_0,\eta\)$ \cite{Riemann-Lebesgue}. For the squeezed configuration we are interested in, $p_1 \approx p_2 \gg p_3$, we have $\(p_1 + p_2 +p_3\)$ very large. Moreover, by definition of the integration limits, $\eta_0\leq\eta'\leq\eta$. Specifically, since $\eta_0$ denotes the time at the beginning of inflation,  $|\eta_0| \gg 1$ whereas $\eta \rightarrow 0$. Therefore, the leading order term from the above integral can be approximated as has been shown above since $1/|\eta_0| \ll 1/|\eta|$. Once we make the above approximation, the equation for the entanglement entropy is given by
\begin{widetext}
\begin{eqnarray}
s_{\rm ent}^{I_1} \sim	\dfrac{(2\pi)^3}{2^7 \Mpl^2}\;\; \int_H^{a H}\dfrac{\d^3p_3}{(2\pi)^3} \int_{aH}^{a\Mpl}\dfrac{\d^3p_2}{(2\pi)^3} \int_{a H}^{a\Mpl} \dfrac{\d^3 p_1}{(2\pi)^3}\;\delta^3(\textbf{p}_1 + \textbf{p}_2 + \textbf{p}_3) \(\frac{p_1 p_2}{p_3^5}\) \; \frac{\epsilon_H\, a^2\, H^4}{\(p_1 + p_2 + p_3\)^2}
\end{eqnarray} 
\end{widetext}
This expression is exactly the same as that in \eqref{I_1} (up to the fact that we have explicitly written down the expression for $\lambda^2$ here and ignored the $\log$ term) and we shall get the same estimate for the entropy of our leading order term. As should be clear from this calculation, all the other terms which have been calculated assuming time-independent perturbation theory retain the same form, up to leading order, even when we relax this assumption and use time-dependent perturbation theory to calculate our matrix elements.

\end{document}